\documentclass[12pt]{spieman}
\usepackage{cite}
\usepackage{amsmath,amsfonts,amssymb}
\usepackage{array}
\usepackage{subcaption}
\usepackage{textcomp}
\usepackage{stfloats}
\usepackage{url}
\usepackage{verbatim}
\usepackage{acro}
\usepackage{multirow}
\usepackage{enumitem}
\usepackage{graphicx}
\usepackage{setspace}
\usepackage{tocloft}
\usepackage[normalem]{ulem}
\usepackage{lineno}
\usepackage{xcolor}
\usepackage{soul}

\DeclareAcronym{sem}{
  short=SEM,
  long=Scanning Electron Microscopy,
}
\DeclareAcronym{tem}{
  short=TEM,
  long=Transmission Electron Microscopy,
}
\DeclareAcronym{vc}{
  short=VC,
  long=Voltage Contrast,
}
\DeclareAcronym{mpsi}{
  short=MPSI,
  long=Multiple Perspective SEM Images,
}
\DeclareAcronym{cd}{
  short=CD,
  long=Critical Dimension,
}
\DeclareAcronym{nl}{
  short=NL,
  long=No Learning,
}
\DeclareAcronym{ml}{
  short=ML,
  long=Machine Learning,
}
\DeclareAcronym{dl}{
  short=DL,
  long=Deep Learning,}
\DeclareAcronym{dnn}{
  short=DNN,
  long=Deep Neural Network,
}
\DeclareAcronym{cnn}{
  short=CNN,
  long=Convolutional Neural Network,
}
\DeclareAcronym{snr}{
  short=SNR,
  long=Signal-to-Noise Ratio
}
\DeclareAcronym{svm}{
  short=SVM,
  long=Support Vector Machine,
}
\DeclareAcronym{dt}{
  short=DT,
  long=Decision Tree,
}
\DeclareAcronym{gpu}{
  short=GPU,
  long=Graphics Processing Unit
}
\DeclareAcronym{cpu}{
  short=CPU,
  long=Central Processing Unit
}
\DeclareAcronym{iou}{
  short=IoU,
  long=Intersection over Union
}
\DeclareAcronym{ap}{
  short=AP,
  long=Average Precision
}
\DeclareAcronym{prisma}{
  short=PRISMA,
  long=Preferred Reporting Elements for Systematic Reviews and Meta-analyses,
}
\DeclareAcronym{tp}{
  short=TP,
  long=True Positive
}
\DeclareAcronym{fp}{
  short=FP,
  long=False Positive
}
\DeclareAcronym{tn}{
  short=TN,
  long=True Negative
}
\DeclareAcronym{fn}{
  short=FN,
  long=False Negative
}
\DeclareAcronym{afm}{
  short=AFM,
  long=Atomic Force Microscopy
}
\DeclareAcronym{rq}{
  short=RQ,
  long=Research Question
}
\DeclareAcronym{hpc}{
  short=HPC,
  long=High Performance Computing
}
\DeclareAcronym{knn}{
  short=kNN,
  long=k-Nearest Neighbor
}
\DeclareAcronym{om}{
  short=OM,
  long=Optical Microscopy
}
\DeclareAcronym{zsad}{
  short=ZSAD,
  long=Zero-Shot Anomaly Detection,
}
\DeclareAcronym{sm}{
  short=SM,
  long=Semiconductor Manufacturing
}

\title{Scanning Electron Microscopy-based Automatic Defect Inspection for Semiconductor Manufacturing: A Systematic Review}

\author[a,b]{Enrique Dehaerne}
\author[b]{Bappaditya Dey}
\author[b]{Victor Blanco}
\author[a]{Jesse Davis}
\affil[a]{KU Leuven, Department of Computer Science, Celestijnenlaan 200A, Leuven, Belgium, 3001}
\affil[b]{IMEC, Angstrom Patterning Department, Kapeldreef 75, Leuven, Belgium, 3001}

\cftpagenumbersoff{figure}
\cftpagenumbersoff{table}

\begin{document}
\maketitle

\begin{abstract}
  In this review, automatic defect inspection algorithms that analyze \ac{sem} images for \ac{sm} are identified, categorized, and discussed. This is a topic of critical importance for the \ac{sm} industry as the continuous shrinking of device patterns has led to increasing defectivity and a greater prevalence of higher-resolution imaging tools such as \ac{sem}. Among others, these aspects threaten to increase costs due to increased inspection time-to-solution and decreased yield. Relevant research papers were systematically identified in four popular publication databases in January 2024. A total of 103 papers were selected after screening for novel contributions relating to automatic \ac{sem} image analysis algorithms for semiconductor defect inspection. These papers were then categorized based on the inspection tasks they addressed, their evaluation metrics, and the type of algorithms used. A notable finding from this categorization is that reference-based defect detection algorithms were the most popular algorithm type until 2020 when \ac{dl}-based inspection algorithms became more popular, especially for defect classification. Furthermore, four broader research questions were discussed to come to the following conclusions: (i) the key components of inspection algorithms are set up, pre-processing, feature extraction, and final prediction; (ii) the maturity of the manufacturing process affects the data availability and required sensitivity of inspection algorithms; (iii) key challenges for these algorithms relate to the desiderata of minimizing time-to-solution which pushes for high imaging throughput, reducing manual input during algorithm setup, and higher processing throughput; and (iv) three promising directions for future work are suggested based on gaps in the reviewed literature that address key remaining limitations.
\end{abstract}


{\noindent \footnotesize Send correspondence to Enrique Dehaerne,  \linkable{enrique.dehaerne@kuleuven.be} \newline Copyright 2025 Society of Photo-Optical Instrumentation Engineers. One print or electronic copy may be made for personal use only. Systematic reproduction and distribution, duplication of any material in this paper for a fee or for commercial purposes, or modification of the content of the paper are prohibited.}

\begin{spacing}{1.5}   

\section{Introduction}\label{sect:introduction}
Advanced semiconductor manufacturing lies at the heart of advancing computer chip technology, aiming to produce more sophisticated and economically viable chips. These chips comprise many interconnected electrical components, notably transistors, forming complex electronic circuits. Central to this endeavor is the drive to minimize the size of these components, historically correlating with significant gains in power, performance, and area \cite{scaling_trends}. Leveraging wafer-manufacturing techniques, wherein multiple chips (also known as dies on wafer) are fabricated concurrently, further underscores the economic advantage of miniaturization. Patterning, the process of creating the structures on the chip that form and connect electrical devices, is therefore the key challenge for advanced semiconductor manufacturing.

\begin{figure}
    \centering
    \includegraphics[width=\textwidth]{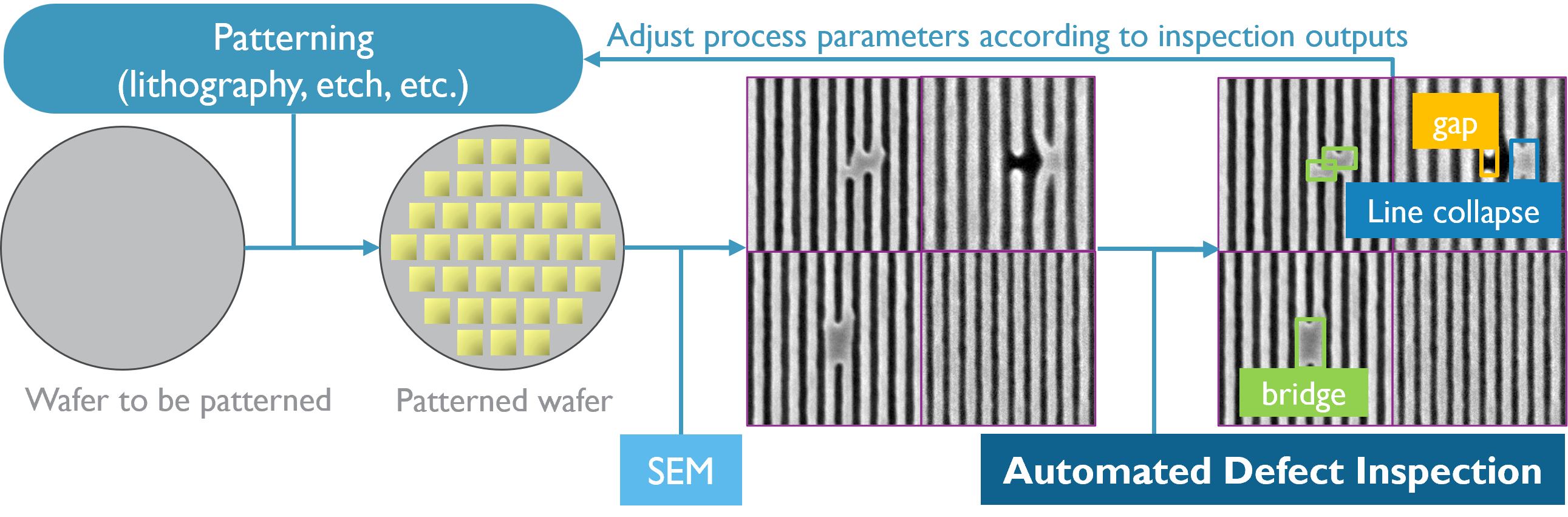}
    \caption{Flowchart showing a simplified overview of the patterning process which relies on defect inspection to guide process optimization.}
    \label{fig:inspection_flow}
\end{figure}

Defects are any errors in patterning or other processes that lower the chip manufacturing yield. High yield must be achieved to make semiconductor manufacturing economically viable. There are many different potential causes for defects including particle contamination, improper process tool parameter settings, or photon-chemical stochastics (becoming increasingly prominent as pattern dimensions shrink \cite{bg_bisschop2019}). Generally, defects are more common in the early stages of process development. Finding and analyzing defects, or defect inspection, is therefore critical to determine the cause of the defects so that the manufacturing process can be refined appropriately to achieve higher yields in subsequent wafers (see Figure \ref{fig:inspection_flow}).

\ac{sem} is very popular for defectivity inspection due to its relatively high resolution, throughput, and versatility. As structures on chips continue to shrink, \ac{sem} is quickly becoming more important relative to \ac{om} thanks to its superior resolution. While \ac{sem} is slower than \ac{om}, it achieves a better trade-off between throughput and resolution compared to other high-resolution imaging methods such as \ac{afm} \cite{holistic_metrology_2011}. Both \ac{sem} and \ac{om} are versatile in that they can be used to measure anywhere on wafers and photomasks. Different imaging modes, most notably \ac{vc} \cite{rw_oberai2017,rw_nakamae2021} and \ac{mpsi} \cite{mpsi_1999}, make \ac{sem} particularly useful in certain scenarios such as buried defects and contrast between different materials, respectively. The combination of all these characteristics is what makes \ac{sem} a critical tool for many inspection jobs, especially at advanced technology nodes.

Algorithms that process and analyze \ac{sem} images enable automation of inspection tasks. This is desirable since manual image analysis is very time-consuming for experts and has even been found to have poor repeatability \cite{sel_Imoto2019, sel_Dehaerne2023b}. Automatic defect inspection promises to relieve experts of the image analysis task, allowing them to focus on optimizing the process using insights from the automated inspection. Research interest in novel inspection algorithms for \ac{sem}-based imaging has increased in recent years. This is thanks to various factors such as mounting talent pressures \cite{kpmg_semicon_industry_outlook_2024}, smaller pattern dimensions that add challenges for conventional algorithms, and the overall increased importance of research in semiconductor manufacturing. Automated defect inspection ultimately improves yield while reducing turnaround time.

This systematic review makes significant contributions to the field by categorizing selected papers and addressing key research questions based on our analysis of them. To achieve this, we first identified relevant papers through a comprehensive search of the four most popular publication databases (IEEE, SPIE, ScienceDirect, AIP) using specific inclusion and exclusion criteria. Through this process, we selected 103 papers for review, which is significantly more than related systematic reviews \cite{rw_derosa2021,rw_lechien2023}. These reviews exclude algorithm types that we include, such as \ac{nl} reference-based algorithms (see Section \ref{subsect:non_learning}). We then categorized the papers based on the inspection tasks they address, the types of algorithms they employ, and the metrics used for algorithm evaluation (Sections \ref{sect:tasks} to \ref{sect:evaluate}). Additionally, we provide a detailed discussion and comparison of the strengths and weaknesses of these categories in their respective sections. Finally, we synthesize new findings from the selected papers to answer the following research questions (Section \ref{sect:discussion_rqs}):
\begin{itemize}
    \item[\textbf{RQ1}] What are the key components for \ac{sem}-based semiconductor defect inspection algorithms?
    \item[\textbf{RQ2}] How does the context of the manufacturing process affect defect inspection algorithms?
    \item[\textbf{RQ3}] What are the key challenges for these algorithms?
    \item[\textbf{RQ4}] What are promising future research directions for the field?
\end{itemize}

\section{Methodology}\label{sect:methodology}
This review used a systematic methodology, following \ac{prisma} guidelines where appropriate \cite{prisma}, for selecting relevant papers. The methodology can be divided into three main steps, each explained further in the following subsections.

\subsection{Relevancy criteria \& search terms}\label{subsect:rel_criteria}
First, a list of criteria was created to determine which algorithms discussed in the papers would be included in the review. These criteria were:
\begin{enumerate}
    \item \textbf{Automatic image analysis:} the algorithm must be able to analyze images without requiring input from human operators. Human control in the form of adjustment of parameters or labeling samples does not violate this criterion as long as it does not have to be done for every image at prediction time.
    \item \textbf{Analysis of \ac{sem} images:} the algorithm must be intended to be used on \ac{sem} images. We also accept variations (for example, \ac{tem}) to the traditional \ac{sem} imaging mode and mention these variations in the discussion of our review when relevant. Papers that experimented with images acquired from other tools such as optical, spectroscopy, acoustic, or physical probing tools were not considered unless they were also explicitly used on \ac{sem} images in the paper. Papers that only used simulated data were also excluded. Other types of data (e.g., design layouts) can be used as long as \ac{sem} images were also used in the paper.
    \item \textbf{Semiconductor defect inspection:} the algorithm must directly produce outputs that determine the defectivity on wafers and/or photomasks. Algorithms for analyzing defects for failure analysis were also considered. Algorithms for other manufacturing quality inspection tasks, such as measurements of \ac{cd}, pattern fidelity, or overlay error, are not considered unless they explicitly enable defectivity analysis in addition to these other tasks. 
    \item \textbf{Novel contribution:} literature reviews and surveys were not considered.
\end{enumerate}

These criteria were then used to create a search command to identify relevant papers from databases. Table \ref{tab:ideal_search_command} shows and explains the search command. The search command was designed to maximize purity, i.e. the number of relevant records retrieved over the total number of records retrieved. Furthermore, an attempt was made to mitigate potential biases towards certain subsets of relevant articles. However, certain search terms had to be changed to optimize the purity of the retrieval. For example, the ``photomask'' search term replaced ``mask'' because ``mask'' retrieved many more non-relevant publications. This could have led to fewer mask inspection papers relative to wafer inspection papers than are representative of the field. Still, the search command reflects our best attempt to retrieve as many relevant articles as possible from the available literature.

\begin{table}[]
    \centering
    \caption{Search command used to find relevant papers to review. Each search term is connected using \textit{AND} operators.}
    \label{tab:ideal_search_command}
    \begin{tabular}{|m{280pt}|m{125pt}|}
        \hline
         \textbf{Search Term} & \textbf{Explanation} \\ \hline\hline
         defect & Domain \\ \hline
         semiconductor OR wafer OR photomask & Relevant domain \\ \hline
         inspection OR review OR classification OR detection OR localization OR segmentation & Relevant inspection tasks \\ \hline
         ``electron microscope'' OR ``electron beam'' OR e-beam OR ebi OR sem & Specify \ac{sem} images \\ \hline
    \end{tabular}
\end{table}

\subsection{Identification of popular publication databases}
Second, popular publication databases were identified by counting the number of papers retrieved from the \textit{Google Scholar} search engine using the search term from Table \ref{tab:ideal_search_command}. Google Scholar was not used to directly search for relevant papers because it retrieved too many results to screen exhaustively (600k+) and has limited search functionality to filter these results. The publishers of the first 150 papers retrieved were counted. The most contributing publishers were SPIE (47), IEEE (37), Elsevier (12), AIP (11), and IOP (9). These were selected for paper identification on their respective search engines except for IOP because of their limited search functionality. All other publishers had 4 or fewer papers retrieved. This search was performed on October 30, 2023.

While this method intuitively identifies publishers that publish many potentially relevant records, it does not consider purity. It is possible that certain publishers do not publish many relevant publications for this review. Consequently, they are not elected as a paper database to be thoroughly searched yet they may have a high purity. Ultimately, the total number of retrieved articles was chosen as the only publisher selection criterion because considering purity at this stage would require additional screening of papers and would risk excluding large publishers that publish many similar but non-relevant papers.

\subsection{Paper identification, screening, \& review}
\begin{figure}
    \centering
    \includegraphics[width=0.9\columnwidth]{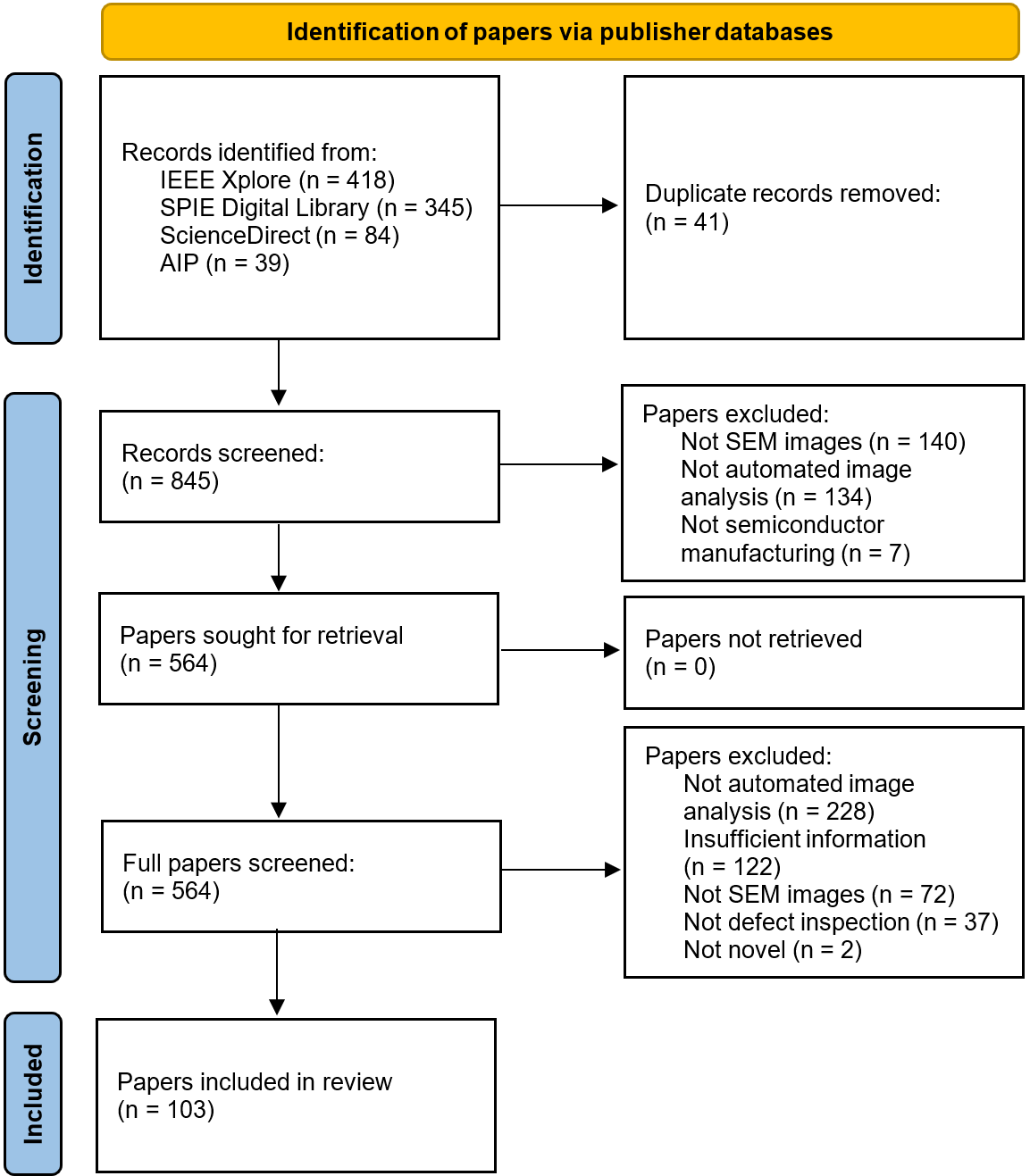}
    \caption{A flowchart of the paper identification \& screening steps followed. The numbers of records/papers included and excluded at each step are also shown.}
    \label{fig:prisma_flowchart}
\end{figure}

Potentially relevant papers were retrieved from the selected publishers' respective search engines and screened to ensure they met the relevancy criteria for this review. Figure \ref{fig:prisma_flowchart} shows a flowchart of this process and the number of records/papers at each step. Note that a record in this case refers to the title and abstract of a full paper, which is either a conference proceedings paper or a journal article. Minor modifications to the search command from Table \ref{tab:ideal_search_command} and additional filters were applied for each search engine where appropriate (see this review's supplementary materials \cite{supplementary_material}). All searches were performed on January 4, 2024. Only records with a publication year of 2023 or earlier were retrieved.

The steps followed to remove irrelevant papers from results retrieved from each search engine were as follows:
\begin{enumerate}
    \item \textbf{Deduplication}: some records seemed to be duplicates, describing the same algorithms but with minor differences in writing or experiments discussed. Records suspected of being near-duplicates were retrieved and cross-examined. If determined to be near-duplicates, only the newest paper of the set was kept. Collectively, 41 near-duplicates were removed.
    \item \textbf{Records screening}: all retrieved records after deduplication were screened for eligibility using the relevancy criteria mentioned in Subsection \ref{subsect:rel_criteria}. If a record did not provide enough information to confidently reject the paper, it was kept for full screening. In total, 845 records were screened and 281 records removed, leaving 564 papers to be retrieved and screened fully.
    \item \textbf{Full papers screening}: papers for all 564 records that passed initial screening were retrieved and screened fully. The papers of all records that passed the initial screening were retrieved and fully screened. Papers that did not provide enough information to recognize the relevancy criteria or describe the method used in sufficient detail were excluded at this stage. Finally, 103 papers made it through the screening process.
\end{enumerate}

All papers selected for inclusion were subsequently reviewed with the categories and research questions introduced at the end of Section \ref{sect:introduction} in mind. For complete data on how each selected paper was categorized, see this review's supplementary materials\cite{supplementary_material}. Note that synthesis of results from the selected papers was not possible due to widely differing applications, datasets, and different metrics reported (if any, see final paragraph of Section \ref{sect:evaluate}).

\section{Defect inspection tasks}\label{sect:tasks}
\begin{figure}
    \centering
    \includegraphics[width=0.75\textwidth]{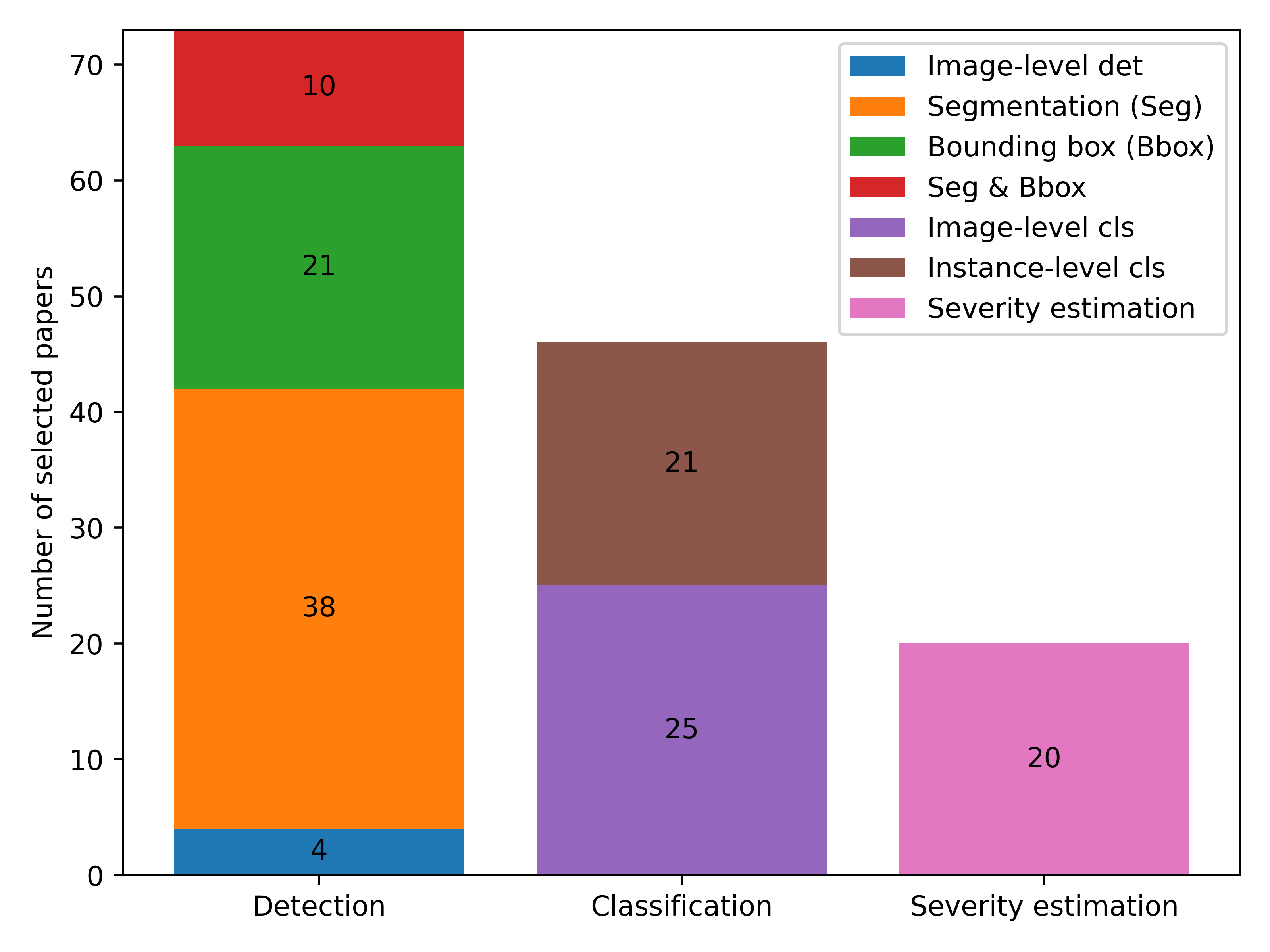}
    \caption{Number of papers that address task (sub-)types.}
    \label{fig:tasks_bar}
\end{figure}

\begin{figure}
    \centering
    \includegraphics[width=0.75\textwidth]{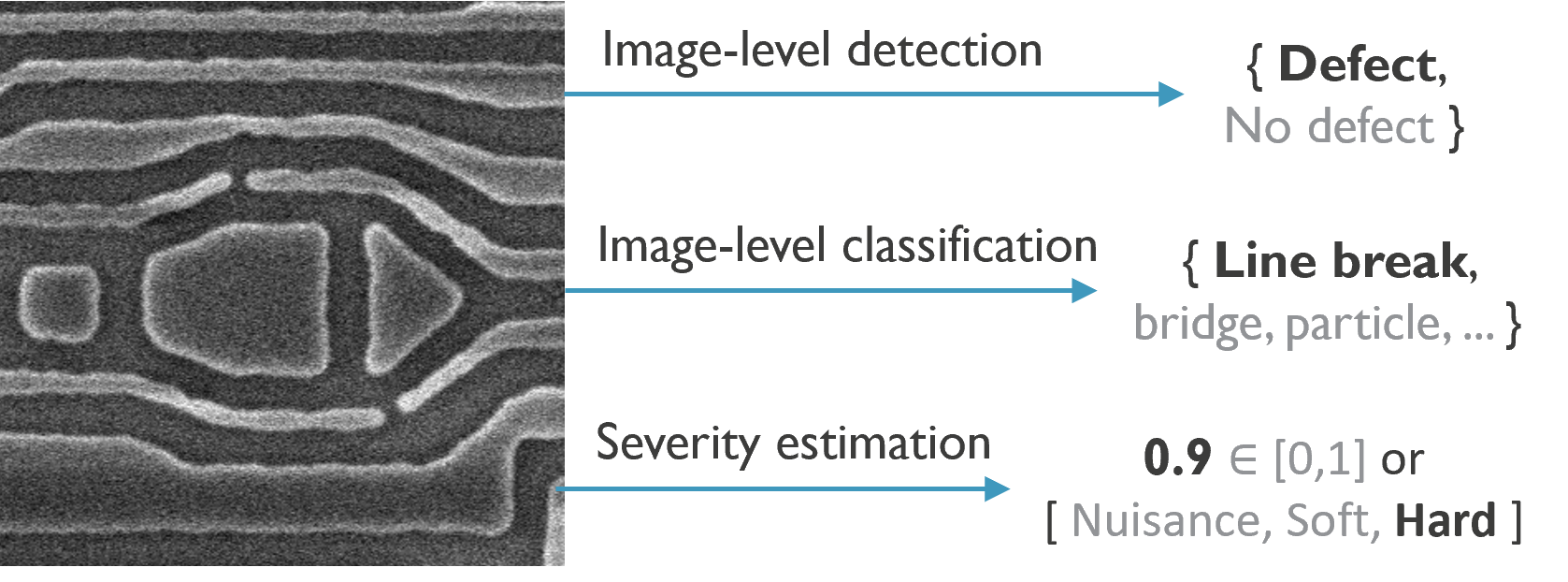}
    \caption{A patterned wafer \ac{sem} image with example predictions for the image-level detection, image-level classification, and severity estimation tasks. The bold black text indicates the correct prediction for the given image, while the other possible predictions are in gray.}
    \label{fig:image_level_tasks}
\end{figure}

\begin{figure}
    \centering
    \includegraphics[width=\textwidth]{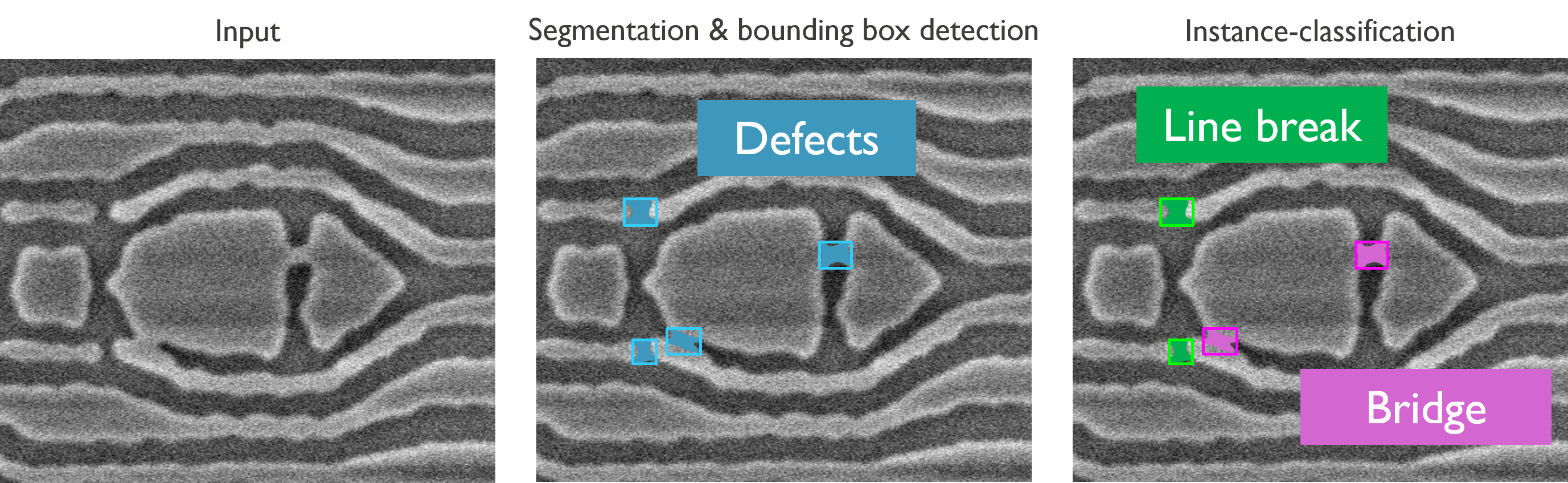}
    \caption{An example \ac{sem} image input (left) with example defect segmentation and bounding box predictions (center). Each of these defect predictions can be classified resulting in an instance-level classification (right).}
    \label{fig:instance_level_tasks}
\end{figure}

Since the terminology and definitions for defect inspection tasks vary between the selected papers, we define these tasks and categorize them. Figure \ref{fig:tasks_bar} shows the number of papers that use algorithms for three main types of tasks and their sub-types. At a high level, defect inspection algorithms are tasked with defect detection (whether there are defects in an image) and classification (what type of defects are in an image). A third, lesser-known task is severity estimation, which estimates how likely it is that a defect will affect yield. These outputs help operators decide how to best mitigate defects in subsequent wafers or photomasks or continue processing the current wafer. Figures \ref{fig:image_level_tasks} and \ref{fig:instance_level_tasks} show example predictions for each main task type and their sub-types. These task (sub-)types are further defined and discussed in the rest of this section.

\subsection{Defect detection}\label{subsect:detection}
Defects are detected in images at either the image or instance level. Image-level detection algorithms take a full \ac{sem} image as input and output a binary signal indicating whether the image contains a defect. This limited output information is generally only useful when assumptions about the input image have been made. This usually includes the assumptions that there is only one defect, often of a particular class \cite{sel_Bonam2013,sel_none2018, sel_lei2022b}, or that the exact number of defects is not important. More generally, images can contain many defect instances and the number and characteristics of each instance may be important. This information can be obtained from localization algorithms that indicate where defects are located in images. The popular types of localization algorithms predict either defect segmentations or bounding boxes.

\subsubsection{Segmentation}\label{subsubsect:segmentation} 
The most popular approach to localizing defects, refers to predicting the location of a defect on the pixel level. In principle, segmentations provide the highest resolution location of defects. Therefore, segmentations are most useful when accurate 2D dimensions for defects are required. However, correctly predicting exactly which pixels in an image belong to defect instances is challenging. For example, determining the precise edges of patterns can be difficult \cite{sel_Gleason2002,sel_Boerger2006,sel_Feng2006,sel_yu2023} and ambiguous \cite{semipointrend}. Additionally, many small pattern variations can resemble actual defects, necessitating the filtering of \ac{fp}, or \textit{nuisance}, predictions on the pixel or instance levels (see Subsection \ref{subsubsect:references}). Despite these challenges, segmenting defects and related tasks such as edge detection are crucial parts of many inspection algorithms, as they provide maximal localization information.

A special type of localization is identifying the location of defects on an image linescan. However, this is not a popular or particularly interesting type of detection due to its heavy reliance on assumptions about pattern structure in the row of pixels being scanned. It will not be discussed further other than noting that the two selected papers that do linescan-based detection \cite{sel_Satya1997, sel_Lin2005} are categorized as segmentation since it is more similar to this task than others.

\subsubsection{Bounding box}\label{subsubsect:boundingbox}
Another way for algorithms to localize defects is by identifying bounding boxes, which are rectangular (or circular in one selected paper\cite{sel_Maruo1997}) patches of an image that contain a defect. Bounding boxes are intended to indicate a patch in the input image where the defect lies but not precisely which pixels in that patch describe the defect. 

Many works start from a segmentation prediction and then predict bounding boxes from it \cite{sel_Lee2014, sel_Toyoda2014, sel_Hsiang2017, sel_Evanschitzky2021, sel_Das2021, sel_Wu2022, sel_Neumann2023}. We assume that this extra step was implemented because it mitigates the more difficult task of predicting defectivity for every pixel. Other possible motivations for this are to connect separated segmentation components that belong to the same defect, which might otherwise be counted as entirely different defect instances, or simplifying downstream processing and analysis. 

Other works analyze images in a patch-wise manner where the image patch is then considered to be a bounding box if a defect is found in it \cite{sel_Maruyama2004, sel_Takeda2008, sel_Graur2015}. This simplifies a complex localization problem into a set of binary image-level detection problems. We should note that these selected papers use fixed patch sizes which are chosen based on assumptions of the size of the defects in the images. This makes these methods difficult to generalize to different scenarios such as varying microscope magnification.

Direct prediction of bounding boxes of variable size would therefore be preferred and is only achieved in the selected papers by certain supervised \ac{dl} algorithms known more broadly as \textit{object-detection} algorithms that learn how to predict bounding-box dimensions from labels provided by human experts (see Section \ref{subsect:deep_learning}). An interesting subset of these algorithms are \textit{instance-segmentation} algorithms, which first predict bounding boxes and then predict the segmentation of the defects within each bounding box, striking an optimal balance between high-resolution localization and being able to distinguish between individual defect instances. 

In conclusion, bounding box prediction is generally most useful for accurately recognizing individual defect instances, which can be tricky to deduce from pixel-level segmentation predictions. However, bounding box localization is not suitable for use cases where pixel-level localization precision is needed unless it is combined with segmentation.

\subsection{Defect classification}\label{subsect:classification} Knowing more about a defect than simply whether it is a defect or not is often crucial for determining how to modify a process to mitigate it. Most defect classification algorithms predict which class a given input belongs to from a predefined set of distinct classes. 

The input can be a full \ac{sem} image or an individual defect instance. For full-image inputs, it is often already assumed that the image has a defect in it. For example, fast throughput tools such as \ac{om} can flag potential defects but cannot classify them due to their low imaging resolution. These potential detected defects are then imaged at higher resolution using \ac{sem} to facilitate classification \cite{sel_Halder2018, sel_Esposito2020}. Instance-level classification is most often performed on individual defect instances obtained from the output of a localization algorithm. Other instance-level classification algorithms perform localization and classification simultaneously, offering advantages such as implementation simplicity and increased processing throughput. A potential disadvantage of this approach is that it restricts the types of defects that can be localized to those belonging to the set of classes the algorithm is programmed to classify\cite{sel_Neumann2023}.

In many classification tasks, the set of possible defect classes has an underlying structure that can be exploited by the classification algorithm. A set of classifiers mimicking a decision tree structure can improve performance when the set of possible defect classes is hierarchical \cite{sel_Ben-Porath1999, sel_Lin2020, sel_Phua2020, sel_Li2022}. An example of this is using an initial classifier to distinguish between broad defect classes such as particle contaminants or pattern structure defects and then selecting a fine-grained classifier, specialized in distinguishing between particle contaminants for example, based on the prediction of the initial  \cite{sel_Ben-Porath1999}. Another possible advantage of this method includes the reusability of coarse-grained classifiers for different classification problems. The main disadvantage of this method is that the setup of many dependent classifiers can be more difficult than creating a single classifier.

\subsection{Defect severity estimation}

Some papers estimate the severity of a defect by assigning it a value that represents either the probability of the defect affecting yield \cite{sel_Mizuno1999} or some related measure such as the resistance added by a defect \cite{sel_Lam2022}. This is similar to classification but differs in that the output is a continuous value or an element in an ordered set which does not directly provide information that could distinguish other characteristics such as the cause of a defect. Aspects that can affect the severity of a defect include contact with patterns \cite{sel_Tomlinson2000} and deviations in expected pattern dimensions \cite{sel_Mizuno1999, sel_Yonekura2005, sel_Kitamura2007, sel_Murakawa2012, sel_Yang2018, sel_Tian2019, sel_Soltani2022} or pixel-value intensities related to material properties \cite{sel_Lagus2005b, sel_Patterson2014, sel_Lam2022}. Two papers discuss algorithms that predict 3D measurements for defects, which we categorize as severity classification algorithms because this type of information is more linked to severity than other task types. It seems to us that, on their own, continuous defect severity estimate values are primarily useful as detailed analysis data for process engineers. Actionable predictions for fully automated inspection flows will most likely require thresholding these estimated values into an ordered set of ordinal predictions that indicate the severity of the problem and give guidance on how to proceed.

\section{Types of inspection algorithms}\label{section:algorithms_learning}
\begin{figure}
  \centering
  \includegraphics[clip,trim={0 150pt 0 150pt},width=0.85\textwidth]{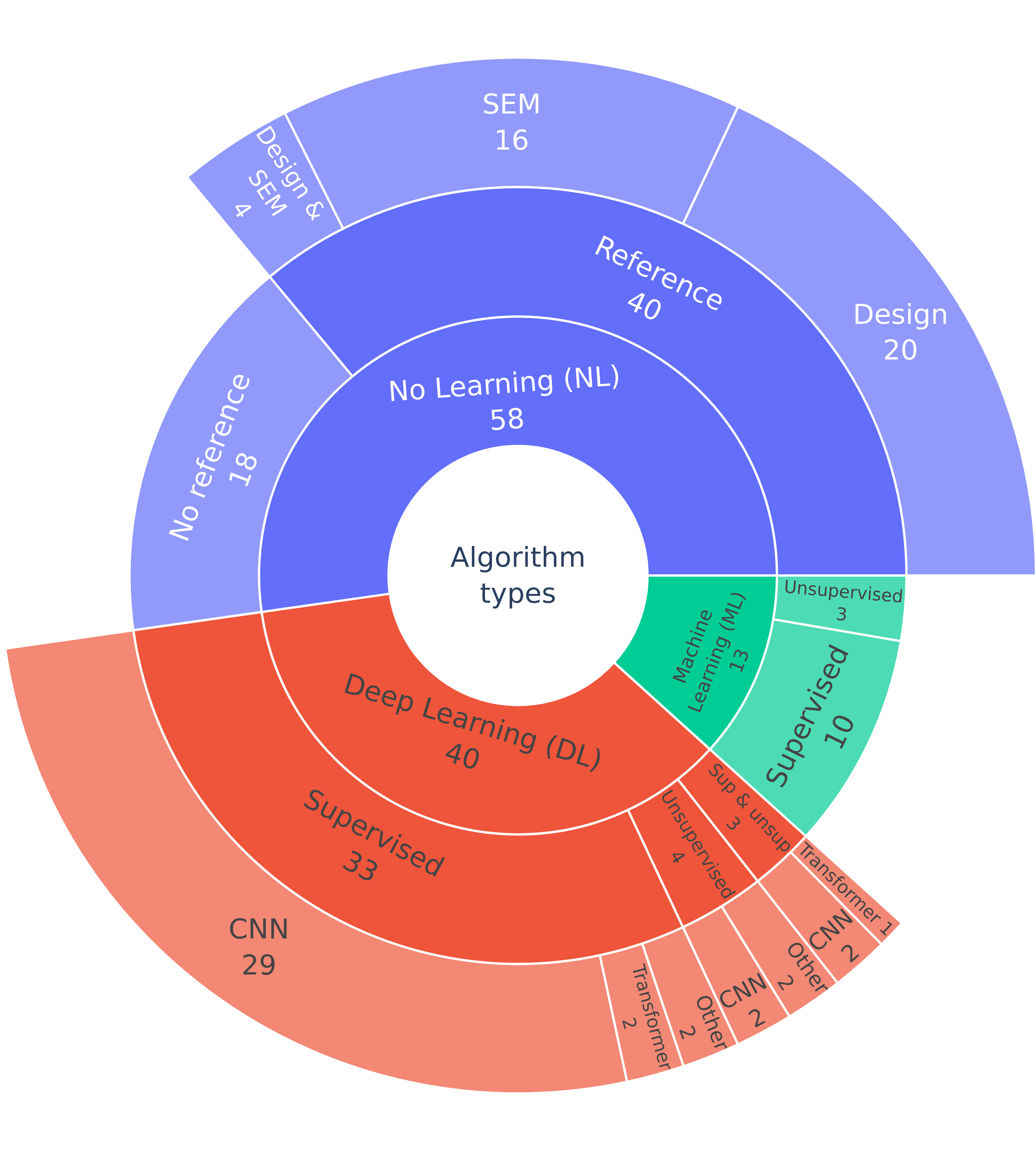}
  \caption{Sunburst chart showing the number of papers that used different algorithm (sub-)types.}
  \label{fig:algos_sunburst}
\end{figure}

Different types of algorithms can be used for each defect inspection task. We categorize these algorithm types primarily by what data they use. The three main types we defined in this sense are \ac{nl}, \ac{ml}, and \ac{dl}. Figure \ref{fig:algos_sunburst} shows the number of papers associated with these types and their sub-types. In general, each algorithm (sub-)type can be used for any previously defined defect inspection tasks but some are better suited for certain tasks. This section discusses these and other factors affecting the performance of inspection algorithms.

\begin{figure}
  \centering
  \includegraphics[clip, trim={5pt 10pt 5pt 5pt}, width=\columnwidth]{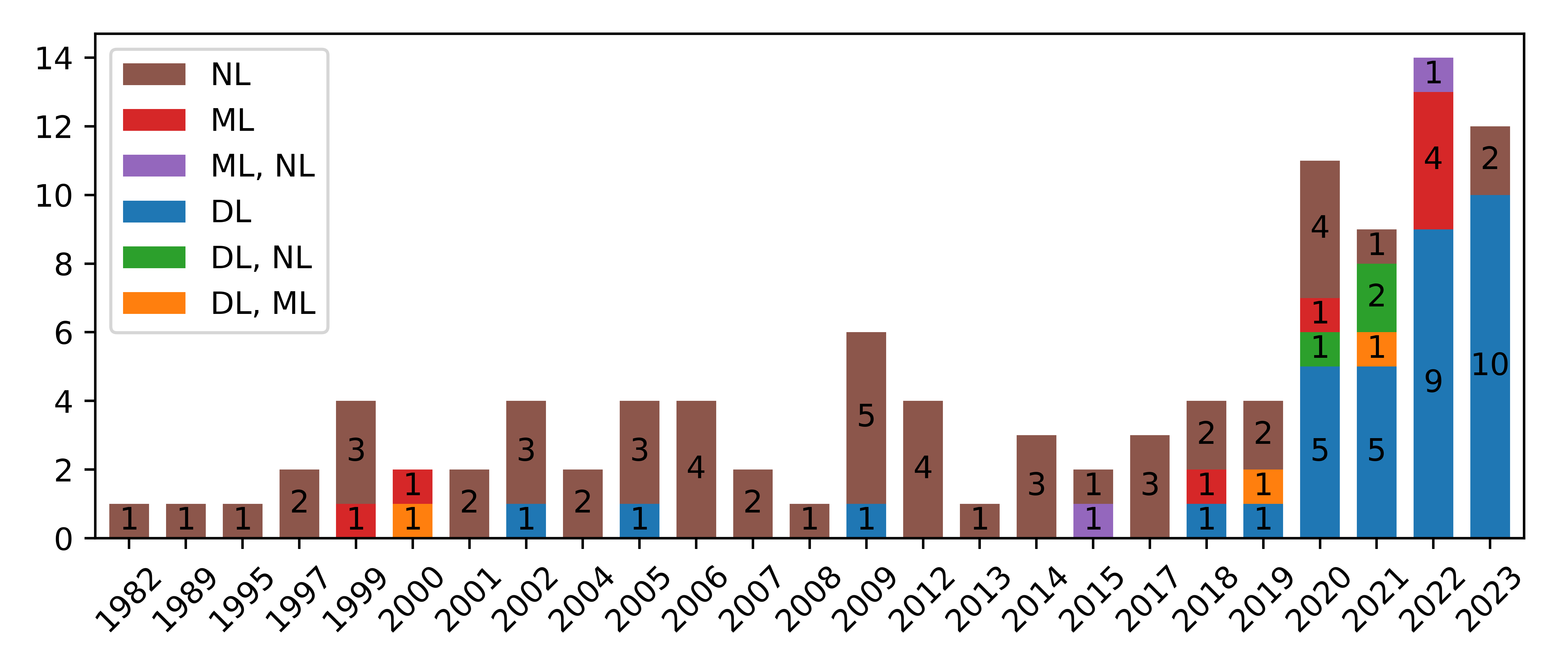}
  \caption{Number of papers per year categorized by algorithm type(s) used.}
  \label{fig:algo_over_years}
\end{figure}

\subsection{\acf{nl}}\label{subsect:non_learning}
\ac{nl} refers to algorithms that are not trained on any data and are instead developed manually. This means that all the parameters of \ac{nl} algorithms must be initialized using expert-knowledge and refined manually through trial-and-error \cite{sel_Okude2021}. This process does not require a representative training dataset which can be a great advantage \cite{sel_Ben-Porath1999}. However, manual refinement can be difficult and time-consuming \cite{sel_Lei2021, sel_Dey2022bb}. In any case, the selected papers seem to suggest that \ac{nl} algorithms were the most popular algorithm category until about 2020 (see Figure \ref{fig:algo_over_years}).

\subsubsection{Reference-based}\label{subsubsect:references}
References provide information about the expected patterns that can be compared to a query image to detect defects. We define reference data as data that does not come directly from the query image (i.e., the sample to be inspected). Additionally, reference data refers to data used directly at prediction time, unlike data used to train \ac{ml} models at setup time.

\begin{figure}
  \centering
  \includegraphics[width=\textwidth]{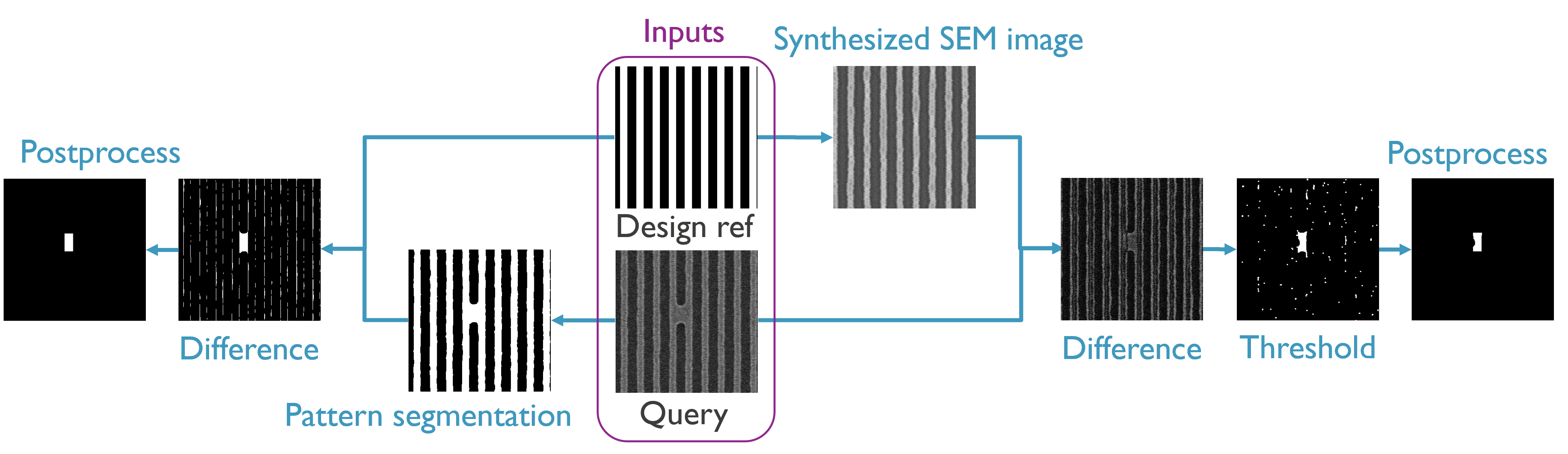}
  \caption{Simplified overview of design reference-based defect inspection. Given an aligned pair of a binary design reference image and a query \ac{sem} image as inputs, a defect segmentation can be obtained by either (left) a difference image of the design and a pattern segmentation from the query or (right) synthesizing a gray-scale \ac{sem} image from the design, taking the absolute difference of the image, and thresholding to get a binary difference image. Postprocessing is usually performed to remove nuisance predictions, morphological transformations were used for postprocessing in these examples.}
  \label{fig:design_comp}
\end{figure}

References can be binary image representations of pattern design information. Segmenting the pattern or extracting pattern edges from a query image and then comparing it with the design data is the most common method of using reference data. The left side of Figure \ref{fig:design_comp} shows a simplified example of this method. Another approach is to synthesize \ac{sem} images from design data using simulators \cite{sel_Endruschat1989, sel_Cho2017} or \acp{dnn} \cite{sel_Okude2021} and compare these synthesized images with the query image. A simplified example of this is shown on the right side of Figure \ref{fig:design_comp}. Both methods leverage the availability of an objective reference of an ideal pattern, provided design data can be aligned with the pattern in the query image.

Comparison with the design primarily enables defect localization, but design reference data can also be used for multiclass classification. Eight papers \cite{sel_Toyoda2014, sel_Hsiang2017, sel_Xie2017, sel_Shah2018, sel_Chen2020, sel_Esposito2020, sel_Okude2021, sel_yu2023} use design information to infer defect classes based on the relative location of defects and target design structures. For example, extra structures could be nuisance defects if they do not interact with the intended structures' electrical characteristics of the target features and this can be determined through the use of design data \cite{sel_Xie2017, sel_Okude2021}. Four of these papers \cite{sel_Toyoda2014, sel_Hsiang2017, sel_Shah2018, sel_Chen2020} specifically mention using multi-layer design data to classify defects in such a way that the root cause can be better traced to a particular process layer.

However, design reference-based methods also have limitations. One main limitation is overcoming the domain gap between the ideal design pattern and a real pattern printed on a wafer. Many wafer pattern features, like corner rounding and edge roughness, are not actual defects but can be mistaken for them because they differ from the target design reference. Some works compensate for this by smoothing design polygons before reference comparison \cite{sel_Graur2015, sel_Xie2017, sel_Chen2020} and/or determining appropriate margins of error for edge differences between the reference and query images \cite{sel_Simpson1982, sel_Maruyama2004, sel_Tsuneoka2006, sel_Kitamura2007, sel_Hagio2009, sel_Toyoda2014, sel_Hsiang2017, sel_Tian2019, sel_Wei2021, sel_yu2023}. \ac{sem} image synthesis from design data can mitigate this but only if the patterning process of the query image is simulated accurately enough. While not mentioned explicitly by these works, another disadvantage of design reference-based methods according to us is the high memory and computational costs associated with processing large volumes of design data. From this, we can infer that these algorithms typically need to be run on remote \ac{hpc} servers. This can extend the time-to-solution (see also Section \ref{subsect:rq3}) and poses data security risks due to the sharing of valuable semiconductor product designs with external \ac{hpc} service providers.

\ac{sem} images can also be used directly as references to mitigate the domain-gap problem from using design reference data. The basic process of \ac{sem} reference defect segmentation is similar to the process of design reference comparison after \ac{sem} synthesis (see the right side of Figure \ref{fig:design_comp}). \ac{sem} images taken from qualified products are excellent references since they include non-defective deviations from the target design, making them more similar to the query image, while not including deviations that are true defects \cite{sel_Yamazaki1999,sel_Hagio2009}. When no references from qualified products are available, \ac{sem} references can be obtained from the same die location on other dies on the wafer (commonly referred to as die-to-die inspection) \cite{sel_Oh2012}. Alternatively, other locations on the same die that have the same target pattern can be used \cite{sel_Okuda2006, sel_Selinidis2009b}. It is usually assumed that references are defect-free, implying that an initial reference selection stage is needed before being able to run the \ac{sem} reference algorithm. Two works \cite{sel_Okuda2006, sel_Oh2012} relax this assumption by averaging many reference images to obtain a reference that is likely to be defect-free even if few of the individual references have defects. Lee and Yoo \cite{sel_Lee2012} used multiple references to improve defect segmentation accuracy. Three papers \cite{sel_Abraham2001,sel_Zontak2009b,sel_Lee2014} use \ac{mpsi}, which captures multiple images from different detectors simultaneously, to improve localization performance by aggregating comparisons from multiple perspectives of the reference and query. In our view, an inherent limitation of \ac{sem} references from non-qualified references is that \textit{systematic} defects that occur frequently over many dies are not detected, as references taken from these dies will also contain these defects. Still, \ac{sem} references can be a valuable alternative to design references especially when design data cannot be used.

The main disadvantage of both design and \ac{sem} reference-based algorithms is that they operate on extremely local features, mostly at the pixel level. Therefore, these algorithms are generally sensitive to small differences between the query and the reference. This includes the differences described above such as pattern roughness but also includes alignment between the query and reference image. While proper alignment is usually assumed as a given by many papers, it is a challenging problem addressed by three of the selected papers \cite{sel_Chon2001, sel_Hiroi2002b, sel_Su2002}. Two papers mitigate the need for perfect alignment by operating on a \textit{kernel} space rather than directly on the pixel space \cite{sel_Zontak2009, sel_Zontak2009b}. These papers mention that many kernel functions are possible. Two papers used \ac{dl}-based methods to use reference data on a learned feature space without direct pixel comparison. We believe this is a promising direction for future reference-based methods (see Section \ref{subsubsect:rq4_multimodal}). In summary, research in \ac{nl} reference-based algorithms focused mostly on pre- and postprocessing of difference images to compensate for the limitations of only considering local features while few works attempted to mitigate this by working on higher-level feature spaces instead.

\subsubsection{No reference}\label{subsubsect:no_ref}
\ac{nl} algorithms that do not use references are more specialized for certain use cases. This also means that the types of tasks they are used for are very diverse. For example, both 3D defect measurement papers use \ac{nl} algorithms without references. Wang et al. \cite{sel_Wang2020b} detect crystal defects in atomic-resolution \ac{tem} images of semiconductor materials using an \ac{nl} algorithm that extracts phases from the periodic structures in the crystals. The separate perspectives captured by \ac{mpsi} are useful for no reference \ac{nl} algorithms to extract particular features from each perspective for defect severity estimation \cite{sel_Ben-Porath1999,sel_Tomlinson2000} or classification \cite{sel_Serulnik2002}. Another relatively popular scenario for \ac{nl} algorithms that do not use references is comparing gray-scale pixel values to expected values. This is most common with \ac{vc} imaging \cite{sel_Satya1997, sel_Lin2005, sel_Matsui2012, sel_Patterson2014} where gray-scale values correlate directly to differences in potential which can be indicative of certain types of defects such as buried defects. However, expected gray-scale values can change significantly between processes and patterns, forcing these methods to \textit{calibrate} to the target dataset. In this case, \ac{ml} or \ac{dl} algorithms would not be a good alternative if there is no representative training data available a priori. This again highlights the specialized nature of no reference methods which only rely on samples of target data and expert-knowledge.

\subsection{\acf{ml}}\label{subsect:machine_learning}
\ac{ml} algorithms are trained on a dataset to make predictions on unseen data. This (partially) mitigates the highly specialized nature of \ac{nl} methods by \textit{learning} expected feature values from the dataset instead of deriving values from expertise and trial-and-error. This confers the advantage that \ac{ml} models can process high-dimensional quantitative data more efficiently than humans, allowing them to find more complex patterns in this data more quickly. Once a dataset has been obtained the \ac{ml} model training is also automatic, meaning that operators are free to spend time on other tasks while it is training. 

However, creating a dataset can require significant effort. Feature extraction is mostly still performed using \ac{nl} methods meaning that many limitations of \ac{nl} methods carry over to the \ac{ml} methods that use them. Some features can be extracted relatively easily from \ac{nl} methods, like feature area and perimeter from segmentations (see left side of Figure \ref{fig:decision_trees}) or brightness in \ac{vc} imaging. However, other, possibly more informative, features can be difficult to extract using \ac{nl} methods, limiting \ac{ml} model performance. Assuming all possible informative features can be extracted, the selection of the most informative features is also non-trivial. Feature selection is often performed manually using domain knowledge but can also be done automatically using so-called feature selection algorithms. Gómez-Sirvent et al. \cite{sel_Gomez_Sirvent2022} showed that finding an optimal subset of features for an \ac{ml} classifier using feature selection algorithms can be computationally expensive. Care should also be taken to ensure the dataset has enough samples to avoid overfitting which is when the model fits the dataset too closely and it does not generalize well to new samples. Still, \ac{ml} algorithms have proven useful for making predictions for defect inspection tasks given a representative dataset with appropriate features.

Many types of \ac{ml} algorithms exist, the \ac{ml} algorithms used by the selected papers can be divided into supervised and unsupervised models. All of them were used for classification or severity estimation tasks. In the rest of this section, we will briefly introduce the important \ac{ml} algorithms used in the selected papers.

\subsubsection{Supervised \ac{ml} algorithms} 
\begin{figure}
  \centering
  \includegraphics[trim={0pt 0pt 0pt 0pt}, width=0.9\columnwidth]{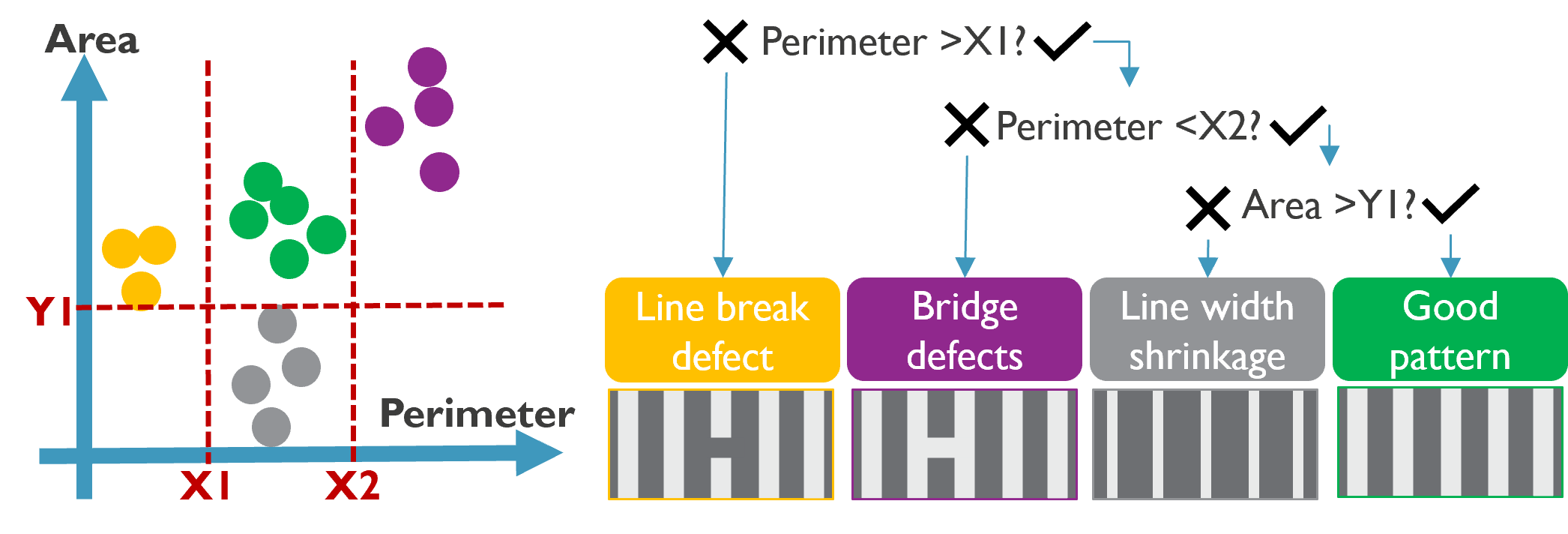}
  \caption{Left: A simple example of a two-dimensional feature space (area and perimeter of the pattern) extracted from images. Right: how a decision tree could partition the space to classify certain defect types.}
  \label{fig:decision_trees}
\end{figure}

\begin{figure}
  \centering
  \includegraphics[width=0.8\columnwidth]{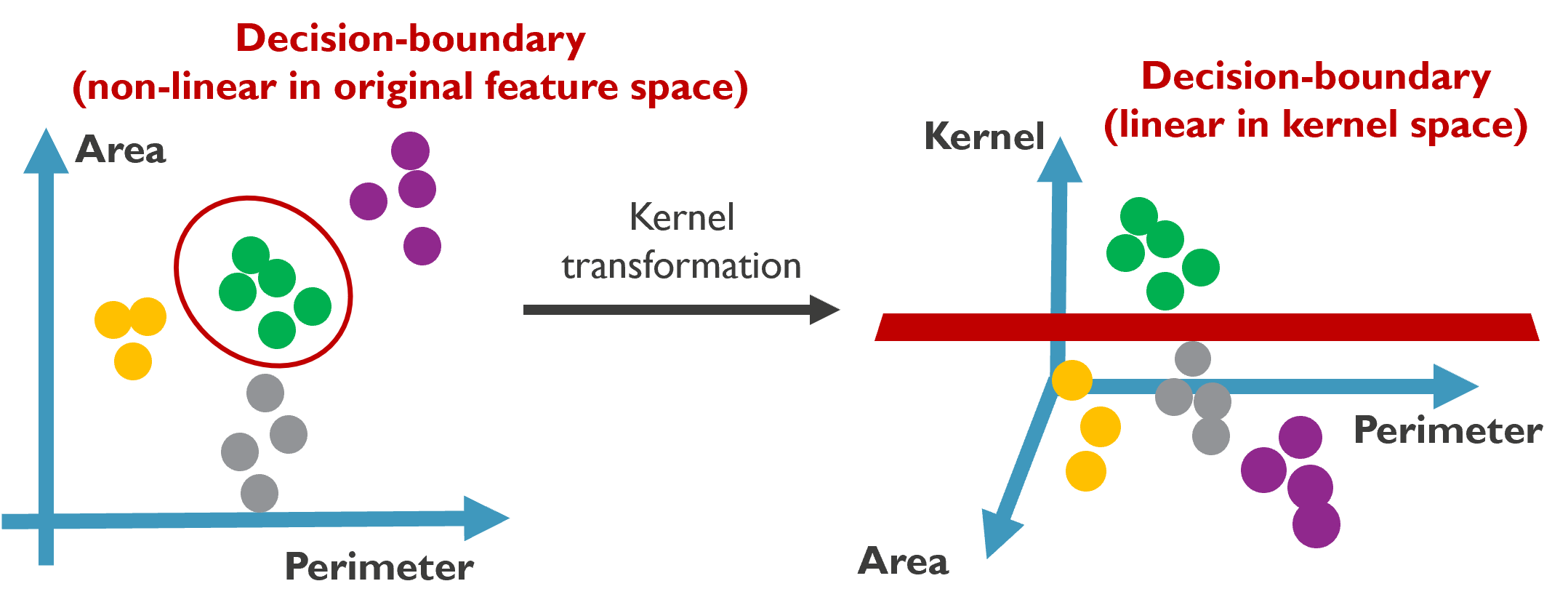}
  \caption{A simple example of binary separation of one group of samples using a hyper-plane defined in the kernel space by an \ac{svm}.}
  \label{fig:svm}
\end{figure}

Supervised \ac{ml} models are trained using pairs of data samples and labels. Supervised \ac{ml} models used in the selected papers include \acp{dt} \cite{sel_Ben-Porath1999, sel_Tomlinson2000, sel_Hunt2000, sel_Lee2020, sel_Lei2022}, \acp{svm} \cite{sel_Graur2015, sel_Lee2020, sel_Gomez_Sirvent2022}, \acp{knn} \cite{sel_Hunt2000, sel_Cheon2019, sel_Arena2021}, and statistical regression (assumed) \cite{sel_Lam2022}. \acp{dt} learn to create rules on the feature space that partition the samples in the training dataset according to their labels, as shown in Figure \ref{fig:decision_trees}. 
\acp{svm} also seek to partition the training dataset but achieve this by defining one or more hyper-planes in a transformed \textit{kernel} space defined by a kernel function.
These hyper-planes in the kernel space effectively define non-linear decision boundaries in the original feature space, as shown in Figure \ref{fig:svm}. These partitions can directly correspond to classes for classification tasks or to continuous output values for severity estimation. 
\ac{knn} algorithms infer labels for an unseen, unlabeled sample based on the distance of the unlabeled sample to the k nearest labeled samples. This means that training a \ac{knn} algorithm simply consists of storing all training sample features and inference consists mostly of calculating distances between the unlabeled sample and all the training samples. While mostly used for classification, distances between samples or partition boundaries can also be used for severity estimation \cite{sel_Tomlinson2000}. With enough training data samples with relevant features, all these \ac{ml} algorithms should learn a mapping between data samples and corresponding labels for unseen samples.

\subsubsection{Unsupervised \ac{ml} algorithms} 
Unsupervised \ac{ml} aims to make predictions on data samples without labels, which is advantageous since labels can be difficult or time-consuming to obtain. All three unsupervised \ac{ml} papers used the k-means clustering algorithm \cite{sel_Halder2018, sel_Soltani2022, sel_Lee2022}. Clustering involves defining a distance metric between features extracted from data samples to indicate how (dis-)similar samples are. These distances are used to define groups or clusters of samples with similar characteristics. These papers use clustering for classification \cite{sel_Soltani2022, sel_Lee2022} or severity estimation \cite{sel_Halder2018, sel_Soltani2022} tasks based on belonging or distance to a cluster, respectively. A disadvantage of these clusters being fully unsupervised is that the clusters they define can be irrelevant to the particular use case they are being used for. With proper feature selection and distance metrics, clustering can be useful when no labels are available.

\subsection{\acf{dl}}\label{subsect:deep_learning}
\ac{dl} is a subset of \ac{ml} that uses multi-layer neural networks, known as \acp{dnn}. For most \acp{dnn}, the input is simply a \ac{sem} image rather than features extracted from the image like with other \ac{ml} models. This is possible due to the automatic hierarchical feature extraction capability of \acp{dnn} that emerges through the model's inner layers before a final layer makes a task-specific prediction. With sufficient training data, \acp{dnn} can make accurate predictions without manual feature engineering.

\begin{figure}
  \centering
  \includegraphics[width=\textwidth]{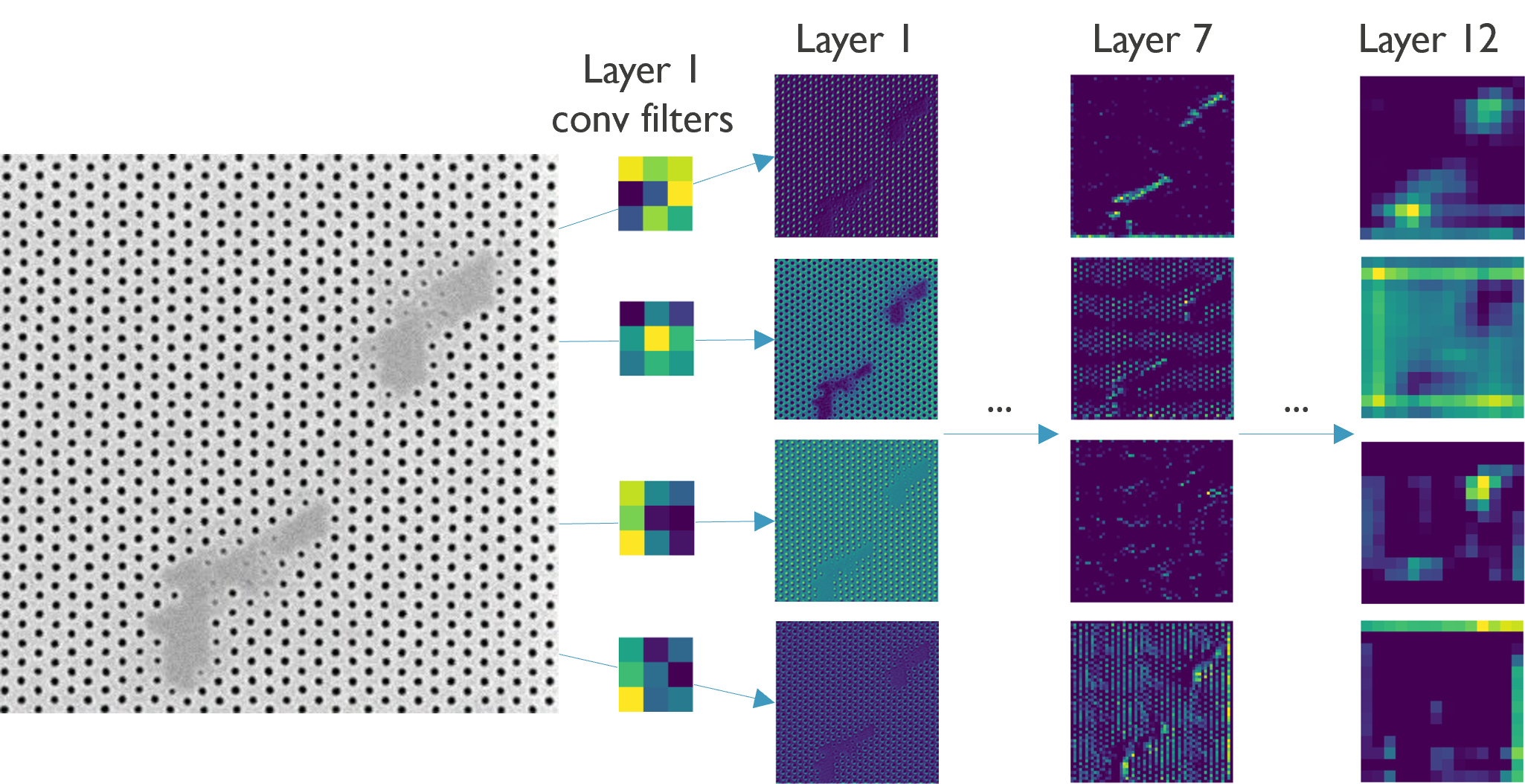}
  \caption{A \ac{sem} image with feature maps output from select convolutional layers from a pretrained VGG16 model\cite{vgg16}. The learned filters of the pretrained model for the first layer of the model are also shown. Each feature map is the result of applying learned convolutional filters over feature maps from previous layers (or the input image in the case of the first layer).}
  \label{fig:cnn_vis}
\end{figure}

Figure \ref{fig:algos_sunburst} shows that the most popular type of \acp{dnn} is \acp{cnn}. \acp{cnn} are specifically designed to exploit characteristics of images, such as relations between neighboring pixels or translation invariance, which makes them more efficient at learning to extract useful image features. More concretely, \acp{cnn} consist primarily of convolutional layers that learn convolutional filters that slide across images to extract feature maps. Feature maps are often downsampled further by a pooling layer. Figure \ref{fig:cnn_vis} shows feature maps extracted from a \ac{sem} image using a prototypical \ac{cnn} model. The early layers extract low-level features such as edges, while later layers extract increasingly more abstract features. 

The efficient feature extraction capabilities of advanced \acp{cnn} have made them so popular for defect inspection that in 2020 they surpassed \ac{nl} as the most widely used algorithm category by the selected papers (see Figure \ref{fig:algo_over_years}). We suggest that this occurred because \acp{cnn} reached a tipping point where the effort needed to curate and label data for training a viable model in a controlled experiment became less than the effort required to handcraft \ac{nl} extraction and prediction algorithms to achieve equivalent accuracy. This data-driven approach could also allow the user to refine their algorithms more easily over time by adding new data samples to the training dataset, for example, rather than manually adjusting algorithm parameters or components. However, in our experience, this popularity has yet to translate to real \ac{sm} environments. We believe that a reason for this is a lack of trust due to the high-dimensional features extracted by \ac{dl} models, which are not easily interpretable by humans\cite{sel_Ben-Porath1999}. This is particularly true when the target images are not similar to images in the dataset used to measure the performance of the model. \ac{nl} algorithms, on the other hand, are inherently interpretable since they were entirely developed by humans. Other, more well studied disadvantages of \ac{dl} relate to practical difficulties with creating a good training dataset. These difficulties will be discussed more in the upcoming subsections. In any case, \acp{cnn} have garnered significant interest in \ac{dl} within the \ac{sm} defect inspection community.

Transformers, a relatively new \ac{dnn} architecture \cite{vaswani2017attention}, have also been used by three selected studies from 2023 \cite{sel_lu2023,sel_Dehaerne2023b,sel_Ridder2023b}. Compared to \acp{cnn}, transformers are more computationally efficient in training and capturing global information but scale poorly with image size and generally require more data to achieve the same performance \cite{transformers_review, dosovitskiy2021vit}. Evidently, implementing and adapting advanced \ac{dl}-based methods is a popular research direction for \ac{sem}-based defect inspection.

\subsubsection{Supervised \ac{dl} algorithms}\label{subsubsect:supervised_dl}
Supervised \ac{dl} models are trained using pairs of data samples and labels, similarly to supervised \ac{ml}. Supervised \ac{dl} is most popular for classification which intuitively makes sense since accurate classification usually requires highly abstract features that are difficult to extract manually for \ac{ml} algorithms but are learned automatically by \acp{dnn}. To facilitate this learning, experts label examples for each class. The main disadvantage of \ac{dl} is that it generally requires larger training datasets than \ac{ml} models. This is limiting in low-data regimes or where labeling costs are too high. The following paragraphs discuss techniques used by selected papers to reduce the number of (labeled) training examples required. We finish this discussion by also discussing label quality.

Data augmentation, modifying data samples to create new ones, is a popular technique used by 19 of the selected papers \cite{sel_Cheon2019, sel_OLeary2020, sel_Phua2020, sel_Fujishiro2020, sel_Patel2020, sel_Yuan_Fu2020, sel_Das2021, sel_Wang2021, sel_Wu2022, sel_Ofir2022, sel_Li2022, sel_none2022, sel_Dey2022b, sel_Dey2022, sel_Dey2022bb, sel_Dehaerne2023, sel_Kim2023, sel_Ridder2023b, sel_Kim2023b}. For supervised learning, data augmentation can reduce the need to label as many data samples. While most of these works use \ac{nl} methods for augmentation, such as random image flipping, two papers \cite{sel_Lei2021, sel_Wang2021} used generative adversarial \acp{cnn} \cite{goodfellow2014generative} to generate new images similar to the training dataset, increasing the number of data samples and improving the performance of their defect classification \acp{cnn}. One selected paper \cite{sel_Wu2022} manually augmented defect-free images by ``painting'' defects onto images. In this case, augmentation was used to address the lack of available data rather than saving human labeling time.

Another popular technique to reduce training data requirements is transfer learning. This refers to taking a \ac{dl} model trained on data from another distribution and training it further on data from the target distribution. Along with generally requiring less data and computation to converge, transfer learning has also been shown to improve generalizability in semiconductor defect detection \cite{sel_Jacob2023}. Five papers specifically mention using transfer learning from natural image datasets \cite{sel_Phua2020, sel_Okude2021, sel_Li2022, sel_Yan2022, sel_Ridder2023b}, three transfer learning from \ac{sem} datasets \cite{sel_Imoto2019, sel_Wu2022, sel_lu2023}, and another three transfer learning from both natural and \ac{sem} datasets \cite{sel_Jacob2023, sel_Ridder2023, sel_Dehaerne2023b}.

Semi-supervised or weakly supervised learning methods use limited labeling time efficiently. Semi-supervised learning involves labeling a limited subset of the training dataset while the rest is unlabeled. This is useful when there is more data available than can be labeled. Kim et al. \cite{sel_Kim2023b} learn from both labeled and unlabeled data simultaneously while Lu et al. \cite{sel_lu2023} first pretrains on unlabeled data in an unsupervised manner and then finetunes on the available labeled data. Both reported improved performance when utilizing unlabeled data compared to only using the limited labeled data. Weak supervision refers to using labels that do not correspond directly with the desired output of the model but are easier/faster to obtain. Patel et al. \cite{sel_Patel2020} use class activation maps \cite{cam_zhou2016} that localize defects in an image while only being trained on image-level defect class labels. Image-level labels are easier to obtain than instance classification labels, allowing a larger dataset to be labeled within the same time budget. Semi-supervised and weakly supervised learning work in different ways to maximize model performance with limited labeling time.

Besides data quantity, labels should also be consistent and as close as possible to the \textit{ground truth to ensure} \ac{dl} model learning converges as intended. Imoto et al. \cite{sel_Imoto2019} found that manual labeling of \ac{sem} images can be noisy and that this data can be used to train models in a pre-training stage, but highly reliable labels should be used for the final, fine-tuning stage of training. Dehaerne et al. \cite{sel_Dehaerne2023} found that inconsistencies in labelers can have detrimental effects on the training of a \ac{cnn}-based defect localization and classification model. They propose a semi-automated method for aggregating labels from multiple labelers to create a more consistent training dataset. Both papers show that label quality should not be overlooked when building a training dataset.

\subsubsection{Unsupervised \ac{dl}} 
Unsupervised \ac{dl} can be split into two main groups, reconstruction-based learning and heuristics-based learning.

Learning to reconstruct input images with incomplete information allows models to reconstruct certain versions of the input image or extract image features without labels. Two papers \cite{sel_Neumann2023, sel_Lee2023} used \acp{dnn} with a bottleneck structure, forcing the model to learn an efficient feature encoding at the bottleneck layer. When trained exclusively on defect-free images, these models fail to reconstruct defects during inference. This allows the model to generate defect-free versions of the input aligned with the input image which can be used as a reference for downstream reference-based defect inspection methods. Dey et al. \cite{sel_Dey2022bb} train a denoising model by corrupting pixels in an image at the input. This incomplete information makes the model unable to reconstruct pixel-wise independent noise from the input which was found to improve downstream defect inspection performance. Lu et al. \cite{sel_lu2023} trains a feature extractor \ac{dnn} to reconstruct parts of the image that are masked at the input given the rest of the image as context. They use this to pre-train the feature extractor for transfer learning to a downstream supervised learning phase. These papers suggest that reconstruction-based learning can be used for various types of processing useful for defect inspection algorithms.

Other unsupervised \ac{dl} methods use heuristics-based objectives to train models to produce outputs resembling desired labels. Two papers \cite{sel_Chang2005, sel_Chang2009b} use objective functions based on heuristics such as pattern disorder, contrast, and pixel brightness to train their models. Ofir et al. \cite{sel_Ofir2022} trained a model using copy-paste augmentation on defect-free images to create synthetic defects. Although they do not look exactly like real defects, these synthetic defects had enough similar characteristics for the model to learn to localize real defects at inference time.

Unsupervised learning offers advantages over supervised learning: (i) it eliminates the need for costly human labeling, and (ii) it avoids biases and false assumptions that can come with labeled data (as discussed in Section \ref{subsubsect:supervised_dl}). However, with a representative labeled dataset, supervised models are expected to perform better than unsupervised models. Ofir et al. \cite{sel_Ofir2022} validate this in their experiments and we find this to be generally true based on our knowledge of the broader \ac{dl} literature.

\section{Evaluation metrics}\label{sect:evaluate}
\begin{figure}
    \begin{minipage}[c]{0.47\linewidth}
    \includegraphics[trim={1350 0 775 755},clip,width=\textwidth]{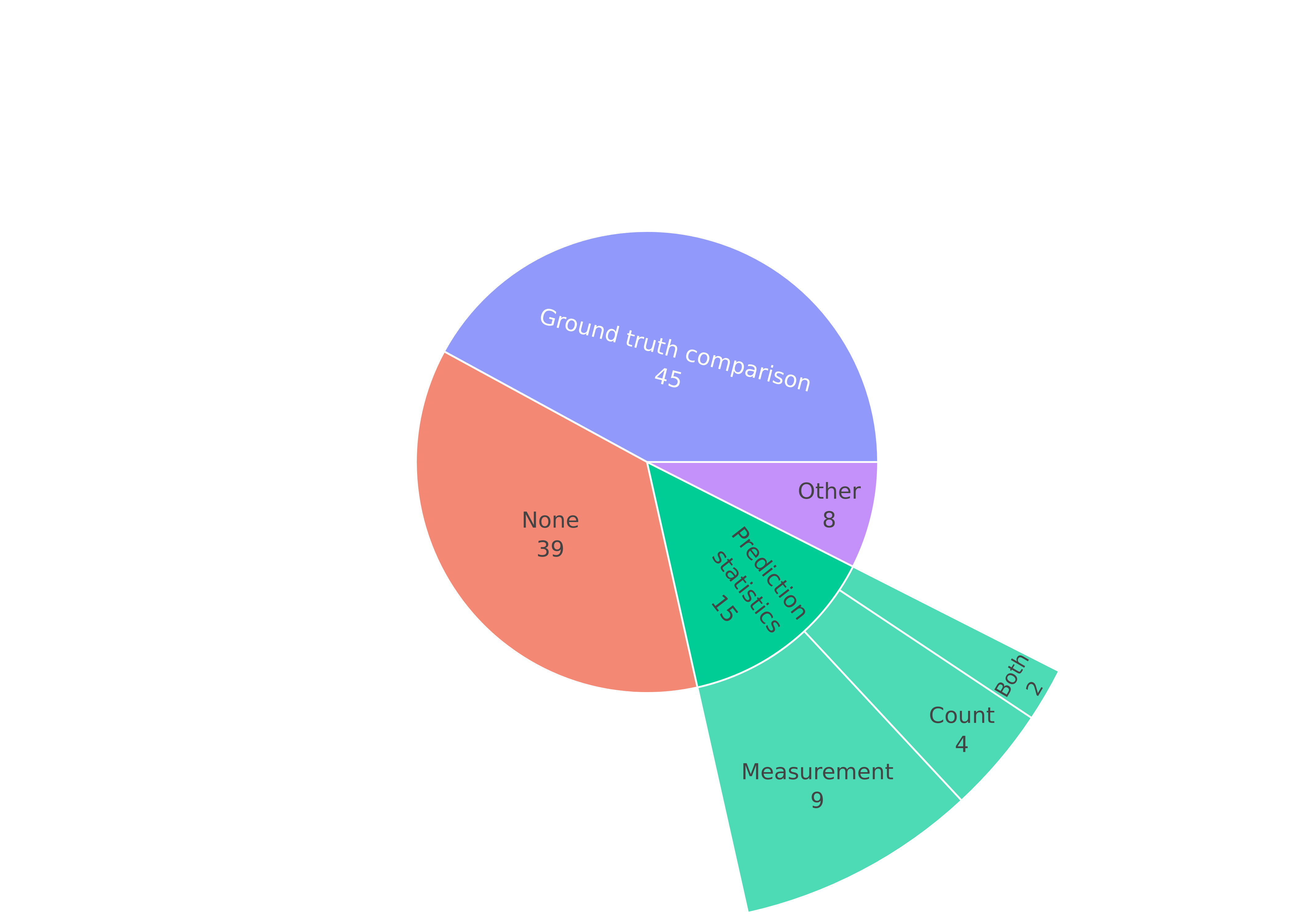}
    \caption{Sunburst chart showing the number of papers that used different metric types. The subtypes for ground truth comaprison-based metrics are shown in Figure \ref{fig:metrics_gt_bars}.}
    \label{fig:metrics_sunburst}
    \end{minipage}
    \hfill
    \begin{minipage}[c]{0.47\linewidth}
    \includegraphics[width=\columnwidth]{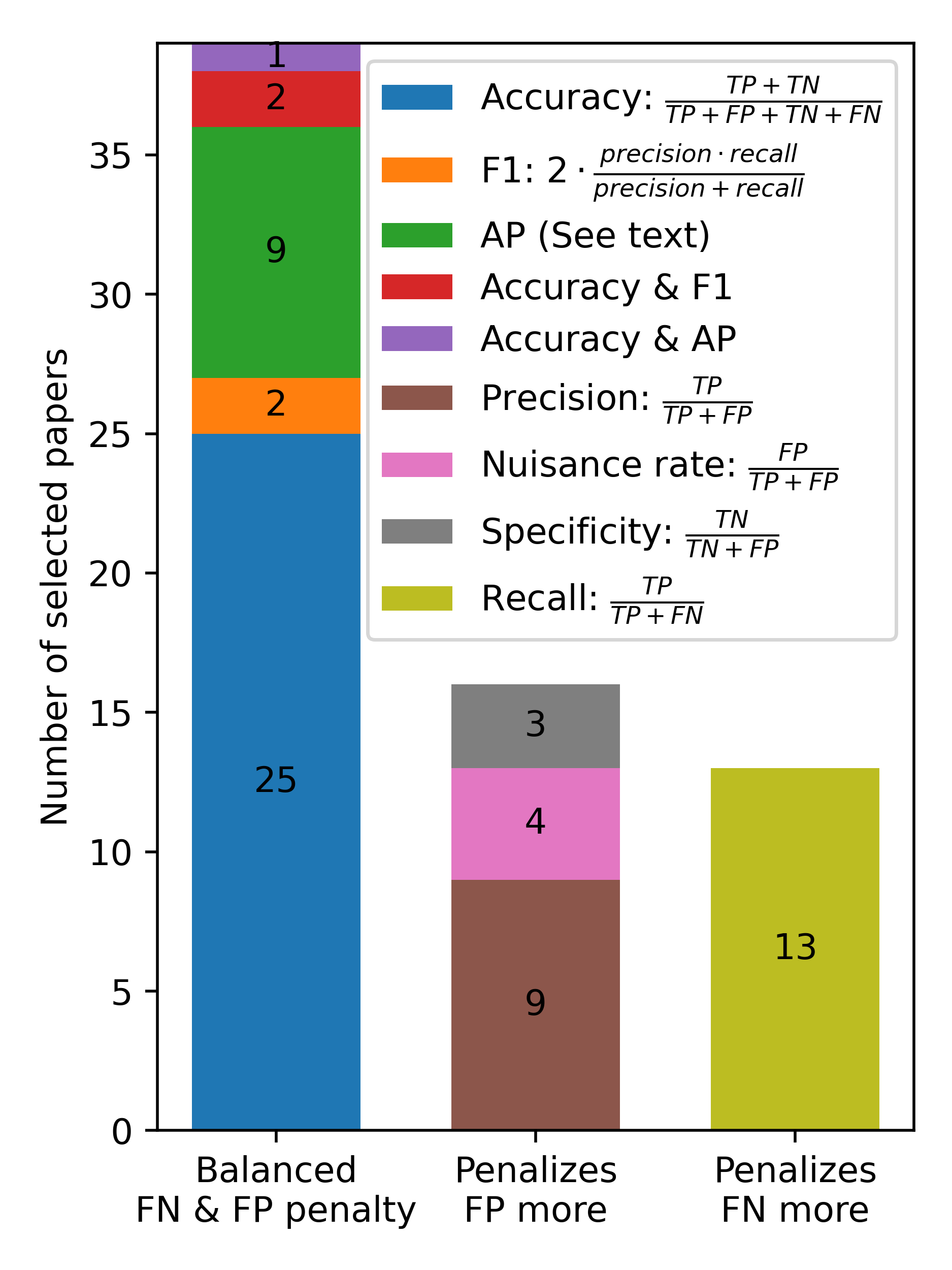}
    \caption{The number of papers that used different ground truth comparison-based metrics.}
    \label{fig:metrics_gt_bars}
    \end{minipage}
\end{figure}

Given a task and target dataset, quantitative metrics are crucial for evaluating the performance of an algorithm and comparing it to other algorithms. As shown in Figure \ref{fig:metrics_gt_bars}, we categorize quantitative metrics primarily by whether they compare predictions to \textit{ground truth} labels.

\subsection{Ground truth comparison metrics}\label{subsect:ground_truth_metrics}
Predictions can be evaluated by categorizing them by the nature of the prediction (true vs. false) and its corresponding ground truth (positive vs. negative). This ground truth is usually obtained from careful manual inspection, similar to how most labeling training datasets for \ac{ml} and \ac{dl} models are labeled. As discussed in Section \ref{subsubsect:supervised_dl}, Dehaerne et al. \cite{sel_Dehaerne2023} found that the ground truth labels assigned through manual inspection can differ significantly between experts, suggesting that even manual inspection should be verified. Additionally, Neumann et al. \cite{sel_Neumann2023} found that their method detected more defects than were initially identified manually, so they manually verified these extra defect detections and recalculated the recall based on these extra verifications as an additional metric. An alternative to manually reviewing \ac{sem} images to obtain ground truths is using additional, more objective tools such as \ac{afm} and electrical testing. Common limitations of these tools are low throughput, destructivity, and tool availability. In general, care should be taken to ensure reliable ground truth data is available for comparison to predictions.

Various methods process the numbers of true/false positives and negatives, defining the different ground truth-based metrics used by the selected papers. Accuracy is the most popular of these metrics, as shown in Figure \ref{fig:metrics_gt_bars}. It measures the ratio of true predictions over all samples. Confusion matrices often accompany accuracy-based evaluation to provide more insights into the prediction behavior of certain classes \cite{sel_lei2022b}. A disadvantage of accuracy is that it does not consider the relative distribution of positive and negative samples. Other metrics such as recall (sensitivity or \ac{tp} rate) and precision (purity) do consider data distribution. Recall and precision should be used when the cost of \ac{fn} and \ac{fp} are high, respectively. Nuisance rate and specificity (\ac{tn} rate) are related to precision because they measure the effect of \acp{fp} (note that a lower nuisance rate is better) but in slightly different ways. Altogether, 16 papers use metrics that penalize \acp{fp} more while 13 papers use recall which penalizes \acp{fn}. This suggests that more papers are interested in reducing \acp{fp} than \acp{fn}. This is in line with the focus of avoiding nuisance defect predictions for many algorithms (see Subsection \ref{subsubsect:references}). The F1 score is defined as the harmonic mean of recall and precision, acting similarly to accuracy in balanced data distributions. The appropriate ground truth comparison-based metric depends on the distribution of positive and negative samples in the dataset and the relative cost of false predictions of both.

\begin{figure}
    \centering
    \includegraphics[width=0.9\columnwidth]{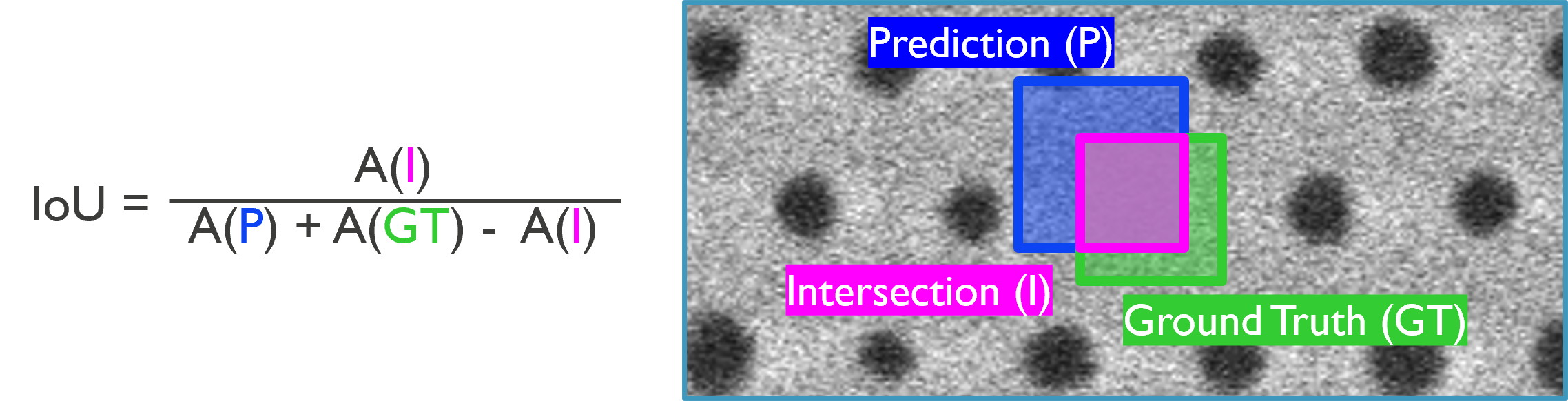}
    \caption{A \ac{sem} image of a missing contact hole defect. Ground truth and predicted bounding boxes are overlayed on the image as well as the box intersecting them. The \ac{iou} formula is shown where $A(X)$ is the area of bounding box X.}
    \label{fig:iou_explanation}
\end{figure}

\ac{ap} is a popular ground truth-based metric for \ac{dl}-based instance localization and classification \cite{sel_kim2021, sel_Dey2022b, sel_Dey2022, sel_Dey2022bb, sel_Ridder2023, sel_Dehaerne2023, sel_Jacob2023, sel_Ridder2023b, sel_Kim2023b}. While all of the previously mentioned metrics can also be used for localization, they often do not use well-defined criteria for when a localization prediction should be considered a \ac{tp}/\ac{fp}/\ac{tn}/\ac{fn} prediction. This is not the case for \ac{ap}. Despite its name, \ac{ap} involves much more than averaging precision values. Given bounding-box or instance segmentation predictions with confidence scores, the steps to calculate an \ac{ap} score are: 
\begin{enumerate}
    \item Calculate the \ac{iou} between predictions and ground truth instances. Figure \ref{fig:iou_explanation} shows how \ac{iou} is calculated from bounding boxes. Segmentation \ac{iou} is calculated similarly and used as used as a metric on its own for one of the selected papers \cite{sel_Gleason2002}. Predictions are considered to be TP if they have an \ac{iou} score above a predetermined threshold and FP otherwise. Ground truth instances not associated with a TP prediction are considered to be FN.
    \item Precision and recall are calculated using the TP and FN values from the previous step. These precision and recall values with associated confidence scores are then used to create a precision-recall curve. This removes the dependency of choosing a confidence threshold for evaluation, although usually predictions with low confidence values are discarded even before \ac{ap} calculation.
    \item The \ac{ap} score is obtained by calculating the area under the precision-recall curve. 
\end{enumerate}
For localization combined with classification, an \ac{ap} score is calculated for every class and can then be combined into a mean \ac{ap} (m\ac{ap}) over all classes. It should be noted that there exists a minimum area under precision-recall that is dependent on the class skew \cite{Boyd2012-ne}. This means that for class imbalanced datasets, which are relatively common in defect inspection use cases \cite{sel_Lee2020, sel_Lei2021, sel_lei2022b, sel_none2022}, the m\ac{ap} can be misleading since the range of possible \ac{ap} values for different classes are not the same.

\subsection{Prediction statistics metrics}\label{subsect:statistics_metrics}
These metrics evaluate an algorithm based on computing aggregated statistics about the values of its predictions (e.g., reporting a histogram of the predicted defect counts and pixel areas\cite{sel_Lagus2005b}). Computing these statistics does not require access to ground truth labels. These metrics can be subcategorized as measurement-based, where the predictions are continuous values most often indicating geometric features, or prediction count-based, where the relative number of discrete predictions of different categories are presented. While these prediction statistics are generally useful for measuring characteristics of the semiconductor products being inspected, they do not directly measure algorithm performance. Only after analyzing the statistics and comparing them to expected values from expert knowledge, for example, can these statistics be used to measure the algorithm's performance. However, we note that this approach is only useful in evaluating the algorithm's performance in aggregate and cannot assess individual predictions. Therefore, we urge researchers to also report ground truth-based metrics by, for example, taking a small subset of the total number of samples being predicted, assigning ground truths to these samples, and using these to compare to the algorithm's predictions. This illustrates that statistics-based metrics are useful only with sufficient context.

\subsection{Papers that do not report quantitative metrics}\label{subsect:no_reported_metrics}
Figure \ref{fig:metrics_sunburst} shows that 40 studies did not report any quantitative evaluation of their discussed algorithms. We believe that this is mainly due to the high levels of competition and confidentiality in the semiconductor industry. We hope that thorough quantitative evaluation of inspection algorithms will become more common in the literature.

\section{Discussion of research questions}\label{sect:discussion_rqs}
This section fully formulates and answers the four \acp{rq} from Section \ref{sect:introduction}. We refer to categorizations and summarize findings from Sections \ref{sect:tasks}, \ref{section:algorithms_learning}, and \ref{sect:evaluate}. Other findings and new ideas are introduced to fill knowledge gaps when appropriate.

\subsection{What are the key components for \ac{sem}-based semiconductor defect inspection algorithms?}\label{subsect:rq1}
With this question, we aimed to identify the common building blocks of all inspection algorithms and how the implementation of each component varies across different algorithm types. We propose that every automated defect inspection algorithm consists of the following sequential phases: (i) setup (ii) pre-processing, (iii) feature extraction, and (iv) final prediction.

\subsubsection{Setup}\label{subsubsect:setup}
Any processing that is done on data other than query images at prediction time we consider as part of the setup phase. A prototypical example is training an \ac{ml} or \ac{dl} model on training data. As discussed in Sections \ref{subsect:machine_learning} and \ref{subsect:deep_learning}, there are various ways of training different models that can be categorized as supervised or unsupervised. Once an appropriate model and training strategy has been selected, most of the remaining work involves curating a training dataset. Two key principles for training data often conflict with available resources. First, the more similar a training dataset is to the target data at prediction time, the better the model's performance will be. This means that models should be retrained on new, more relevant datasets as they become available to achieve maximum performance. However, each retraining may require significant manual dataset curation and labeling along with the computation needed to retrain or fine-tune the model. Second, the larger and more diverse the training dataset, the less the model will overfit to specific examples, improving predictions for unseen data samples. Advanced \ac{dl} models with automatic feature extraction capabilities require particularly large datasets to learn useful and generalizable features. Transfer learning is popular for overcoming this problem. Keeping a feature extraction network frozen, meaning that only the final prediction parts of the model are updated during training, can further improve training convergence time\cite{sel_Li2022}. Overall, the setup phase is particularly important for \ac{ml} and \ac{dl} models that must be trained on curated datasets.

\ac{nl} methods also have setup phases, such as selecting appropriate algorithm parameter values, handcrafting features, or selecting references. Threshold values are particularly sensitive to changing imaging or processing conditions and should be reassessed or redefined in the setup phase. An example of an important threshold value is minimum defect dimensions that will filter out nuisance defect predictions caused by small feature variations. Similarly, edge detection algorithms often rely on carefully selected threshold values for good results \cite{sel_Okude2021}. In some cases, such as classifying new and different defect types, modifying parameter values is not enough, and new feature extraction operations must be added, which can be challenging and time-consuming. Many reference-based \ac{nl} methods also need to set up references before prediction time. This can include design pattern region of interest identification \cite{sel_Xie2017}, reference averaging \cite{sel_Oh2012}, or manually selecting an optimal reference image from a set of candidates. The exact manner and extent of the setup changes to an \ac{nl} algorithm will be determined from pattern design rules or domain knowledge, often necessitating trial-and-error on a representative dataset to finetune appropriately.

All inspection algorithms should also be evaluated before deployment. The selection of the appropriate metric and how the metric is calculated, especially the collection of ground truth data (as discussed in Section \ref{subsect:ground_truth_metrics}), should be considered as an important part of the setup phase. For supervised \ac{ml} and \ac{dl} algorithms, this follows from the labeled data they use and the learning objectives they optimize, but for many unsupervised and \ac{nl} methods, evaluation constitutes a separate setup step.

\subsubsection{Pre-processing}\label{subsubsect:preprocessing}
Raw input images can be transformed into a similar image to improve subsequent feature extraction. While this can include operations such as input data normalization (resizing, data type, etc.), the most consequential types of pre-processing are image enhancement algorithms. Image enhancement includes denoising \cite{sel_Yonekura2005, sel_Boerger2006, sel_Chang2009b, sel_Hsiang2017, sel_Chen2020, sel_Wei2021, sel_Dey2022bb}, contrast enhancement \cite{sel_Yonekura2005, sel_Wei2021}, and distortion correction \cite{sel_Chon2001}. Another type of preprocessing used by a selected paper includes combining images taken from \ac{mpsi} into a single image with higher contrast \cite{sel_Murakawa2012}. These pre-processing operations are usually optional, but can be helpful since they preserve the fundamental characteristics of the image while highlighting signals of interest.

\subsubsection{Feature extraction}\label{subsubsect:feature_extraction}
Extracting machine-interpretable descriptions of images or regions is a critical component of every automatic defect inspection system. \ac{nl} and \ac{ml} (excluding \ac{dl}) methods use handcrafted features. Difference images obtained through alignment, pixel-wise comparison, and thresholding between query and reference images are an example of handcrafted image feature that facilitates defect segmentation. Edges detected using an algorithm such as Canny \cite{canny_edge_detection} may also facilitate defect segmentation. Common classification task features include descriptions of segmented areas such as area, perimeter, or average grayscale value \cite{sel_Lam2022, sel_Lee2022}.

Advanced \acp{dnn} automatically learn to extract hierarchical, complex, and abstract features from images during training, often in an end-to-end manner with a feature classification network. While these learned features are useful for subsequent feature classifiers, they are largely uninterpretable for humans. This is unlike \ac{nl} methods where the feature extraction operations are created, and thus interpretable, by human experts \cite{sel_Ben-Porath1999}. To make the features extracted by their \ac{dnn} more interpretable, one paper \cite{sel_Patel2020} used a special \ac{dnn} architecture, namely class activation maps \cite{cam_zhou2016}, which gives insight into which regions of an image contributed most to a classification output. Another special case is autoencoder \acp{dnn} for reference image generation \cite{sel_Neumann2023, sel_Lee2023}. The generated reference images are interpretable and are usually followed by additional \ac{nl} feature extraction which is also interpretable. Still, a key limitation of the automatic feature extraction capabilities of most \ac{dl}-based inspection methods is its lack of human interpretability of all the features in the many layers of \acp{dnn}.

\subsubsection{Final prediction}\label{subsubsect:final_prediction}
Predicting a task-specific output from the extracted features is the final component of an inspection algorithm. This is rather straightforward for supervised \ac{ml} models which take the features as input to the trained model and produce an output corresponding to the expected label. Similarly, supervised \ac{dl} models use a (part of a) \ac{dnn} to extract image features and then feed them to another (part of the) \ac{dnn} to produce a label prediction. Unsupervised \ac{ml} and \ac{dl} algorithms are different in the sense that the prediction is not associated with a label but rather to (dis-)similarity with the training data \cite{sel_Hunt2000, sel_Halder2018, sel_Cheon2019, sel_Arena2021, sel_Soltani2022, sel_Lee2022, sel_Neumann2023, sel_lu2023, sel_Lee2023} or heuristics \cite{sel_Chang2005, sel_Chang2009b, sel_Ofir2022}. For many \ac{nl} segmentation methods, this is where segmentation maps are refined to exclude nuisance defect predictions \cite{sel_Hiroi2002b, sel_Gleason2002, sel_Lee2012, sel_Kim2015} or connect separated defect segmentation components \cite{sel_Feng2006, sel_Lee2014}. Final prediction is often a relatively simple but critical component of a defect inspection algorithm that must be tuned most for particular inspection jobs.

\subsection{How does the context of the manufacturing process affect defect inspection algorithms?}\label{subsect:rq2}
We propose that the maturity of the manufacturing process, or how far the process has progressed on the yield learning curve \cite{xu_yield_learning_curve_2024}, affects the availability of input data. This is a critical factor for deciding on an appropriate algorithm type, as discussed at length in Section \ref{section:algorithms_learning}. Early in the yield learning curve new devices and process flows are researched and developed. At this stage, not much data has been collected and process parameters are changing relatively frequently which can render previously collected data unrepresentative for subsequent inspections. \ac{ml} and \ac{dl} algorithms are most affected by this since it limits their training datasets and requires frequent retraining to achieve good performance. This is especially true for supervised classification methods when a process change causes a new type of defect, which the model cannot predict properly without retraining on data with these new defects. Similarly, for reference-based algorithms, a defect-free reference will be harder to find and will need to be updated as the process evolves. Generally, all inspection methods become more viable and performant as the process matures towards high-volume manufacturing, with more available data and less variation between subsequent inspection jobs on the same process.

We also propose that the output requirements of defect inspection algorithms are significantly influenced by the maturity of the process. In the early stages of the yield learning curve, there is a higher number and greater severity of defects. Consequently, inspection efforts need to focus on detecting and classifying \textit{hard} defects, which are anticipated to have a substantial impact on yield \cite{sel_kim2021}. In later stages, the focus shifts to \textit{soft} defects, which may or may not affect device performance \cite{sel_kim2021}, as hard defects become increasingly rare and process parameters are fine-tuned rather than undergoing drastic changes that could cause new types of defects. For \ac{nl} methods, adjusting detection and classification threshold values is crucial for targeting specific types of defects. Tuning of \ac{ml} and \ac{dl} algorithms usually involves curating the training dataset, especially its labels, to reflect the desired outputs, followed by retraining the models. In summary, as the yield ramps up, the overall defectivity decreases and the need for corresponding process changes diminishes, impacting both the setup phase of the algorithms and their expected performance.

\subsection{What are the key challenges for these algorithms?}\label{subsect:rq3}
We propose that minimizing time-to-solution, the total time needed to complete a defect inspection job from imaging to output prediction analysis, while delivering minimally viable performance, is the key challenge of a defect inspection system. The inspection system can theoretically be optimized to perfectly fit a specific use case but will not work on a similar but different use case, prompting a long algorithm optimization process all over again. On the other hand, a poor-performing inspection system will require extra work, often in the form of manual image inspection, to correct false predictions which adds substantially to the overall processing time. Note that the performance level that could be considered minimally viable is often high but varies depending on the exact scenario. The performance of defect inspection algorithms is therefore of utmost importance but has already been discussed in this review. In the rest of this subsection, we focus on the challenges relating specifically to the time aspect of the time-to-solution problem.

\subsubsection{Acceptable performance with high imaging throughput}
Often, the bottleneck for metrology of advanced nodes is \ac{sem} imaging due to its relatively low throughput and high demand for many use cases. A tradeoff exists between the imaging throughput and the \ac{snr} of the images obtained. Acquiring more frames of a sample reduces noise in the average but requires more electrons to be fired at the sample, which takes more time. High \ac{snr} generally gives better performance for defect inspection algorithms so, ultimately, imaging recipes should be tuned to balance between imaging throughput and inspection algorithm performance \cite{sel_Takeda2008}. Using algorithms that are more robust to low \ac{snr} allows for higher throughput imaging. Two papers \cite{sel_Das2021, sel_Dey2022bb} mention that \ac{dl} methods are more robust to noise than \ac{nl} methods while Neumann et al. \cite{sel_Neumann2023} explicitly include data of various noise levels in their training set to make their \ac{cnn} more robust to noise. Okuda and Hiroi \cite{sel_Okuda2006} and Oh et al. \cite{sel_Oh2012} use averaging of aligned images to capture a \textit{golden} reference image for defect comparison while being able to image every individual image at high throughput rates. Image enhancement algorithms (see Subsection \ref{subsubsect:preprocessing}) are another way to maintain defect detection performance while relaxing image quality requirements, allowing for higher throughput imaging. Any algorithmic techniques that make the tradeoff between inspection performance and imaging throughput more favorable can significantly improve total time-to-solution.

\subsubsection{Manual input during setup}
The setup phase in the algorithm deployment cycle (see Subsection \ref{subsubsect:setup}) often requires manual input. This manual work is especially expensive due to talent limitations in the semiconductor manufacturing industry \cite{kpmg_semicon_industry_outlook_2024}. Ways to (partially) automate the setup of \ac{ml} and \ac{dl} models include automatic feature selection (mainly for non-\ac{dl} models, see Section \ref{subsect:machine_learning}) and semi-supervised learning (see Section \ref{subsubsect:supervised_dl}). Although not discussed in any of the selected papers, we believe that having a labeled evaluation dataset for \ac{nl} methods could allow for simple brute-force or learning algorithms to predict initial parameter values, potentially speeding up the manual tuning process. More generally, labeled evaluation datasets can often be (partially) reused throughout the manufacturing optimization process, enabling more repeatable and faster evaluations in subsequent inspections. This contrasts with manual qualitative evaluations, which might be quicker in the short term but slower in the long term due to having to perform many iterations.

\subsubsection{Processing throughput}\label{subsubsect:throughput} 
Processing throughput is the time needed to predict defectivity for all given images. An obvious way to improve this is to use or design algorithms that can analyze images quickly, as many selected papers claim \cite{sel_Hiroi2002b, sel_Serulnik2002, sel_Lin2005, sel_Feng2006, sel_Zontak2009, sel_Lee2012, sel_Yang2018, sel_kim2021, sel_Yan2022, sel_Dey2022, sel_Ridder2023b, sel_Ridder2023, sel_Dehaerne2023}. These design choices are quite specific to each algorithm. Ultimately, processing throughput is determined by the type and amount of available computational resources. 

\acp{gpu} specialize in matrix multiplication operations and parallelization, greatly speed up computation in many cases. \acp{dnn} are especially well suited to take advantage of these capabilities. At least an order of magnitude improvement in throughput on \ac{gpu} compared to \ac{cpu} was reported in two selected papers \cite{sel_Imoto2019, sel_none2022}. One \ac{nl} method \cite{sel_Zontak2009} was implemented for improved throughput on \acp{gpu} and another \cite{sel_Lee2012} proposed a \ac{gpu} implementation as future work to improve throughput. While not mentioned in the selected papers, many types of \ac{ml} models have implementations that are designed for \acp{gpu} which can be especially beneficial for speeding up training \cite{catboost_gpu,svm_gpu,kmeans_gpu}.

Many \acp{cpu} and \acp{gpu} can be used together in parallel along with significant memory and storage resources in an \ac{hpc} platform for maximum throughput and utilization. An example of particularly memory-intensive algorithms that generally need to be run on \acp{hpc} are design reference-based inspection algorithms, also referred to as die-to-database algorithms \cite{sel_Boerger2006, sel_Tsuneoka2006, sel_Kitamura2007, sel_Hagio2009, sel_Yang2018}. These algorithms process large files of design data, including pattern structure geometries and associated design rules, and batches of \ac{sem} images, often including their imaging metadata, for inspection. Additionally, many die-to-database algorithms also make use of computationally intensive lithography simulators for synthesizing \ac{sem} references \cite{sel_Endruschat1989, sel_Cho2017} or estimating printed patterns on wafer from extracted contours on photomasks \cite{sel_Yamanaka2004, sel_Guo2009, sel_Murakawa2012, sel_Cho2017}. These simulators should also be run on \acp{hpc} to keep execution times reasonable. Investment into acquiring and managing compute and memory resources is a prerequisite for running advanced defect inspection algorithms at high throughput.

\subsection{What are promising future research directions for the field?}\label{subsect:rq4}
To address significant gaps in the literature, we suggest the following three research directions for the defect inspection algorithm research community.

\subsubsection{Generalizable inspection algorithms}
Defect inspection algorithms that are generally applicable to diverse and challenging pattern types would skip or automate many setup steps that must currently be done manually. These algorithms would enable automatic inspection when no or very small representative datasets are available and design data cannot be used, a scenario which no current algorithm is reliable without significant setup effort. Methods from selected papers that come close to realizing this include reference averaging\cite{sel_Okuda2006, sel_Oh2012} and non-reconstruction-based unsupervised learning\cite{sel_Chang2005, sel_Chang2009b, sel_Ofir2022}. However, these methods make assumptions that limit their applicability to simple, repeated pattern types in their current form. While further investigation into advanced techniques discussed in the selected papers such as self-supervised pretraining, semi-supervised learning, and verifying label quality would make for interesting research, we encourage researchers to investigate emerging \ac{dl} techniques not yet been applied to \ac{sm} defect inspection. \ac{dl}-based \ac{zsad}, which aims to detect \textit{anomalies} (defects in the case of semiconductor defect inspection) in a target image without having seen a normal or abnormal target example, is a promising related field for generalized detection. Batch-wise \ac{zsad} \cite{zsad_batch_norm_2023,li2024musc}, where a batch of target images are analyzed together at prediction time, is promising for exploiting the availability of multiple images of the same pattern from the same wafer to improve \ac{zsad} performance. In essence, these methods work on the same principles as average image referencing \cite{sel_Okuda2006, sel_Oh2012} without the strict reference image alignment requirements. For classification, deep image feature clustering methods \cite{vangansbeke_scan_2020} show promise in identifying groups of semantically similar images which can then be used to quickly classify many instances from few labels. Other approaches, such as meta-learning, where models are trained to learn new tasks quickly, are promising for initializing models that can learn from few examples efficiently \cite{fsl_survey}. We should note that these emerging methods are unlikely to reach the same level of inspection performance compared to the \ac{dl} methods already investigated by the selected papers when reasonable amounts of representative data are available. These data-efficient methods show promise as a first solution early in the process learning curve and can possibly supplement established methods later in the curve to recognize new defects from process drifts due to their generalizability.

\subsubsection{Automated microscopy for real-time inspection algorithms} Currently, the \ac{sem} imaging process is controlled by fixed, predetermined \textit{recipes}. While this is a simple way to program a \ac{sem} for an inspection job, it does not allow for feedback from the image to change the imaging parameters as needed. An example of when this could be useful is if, after only a few image frames, the class of a defect is already clear to a classification algorithm. In this case, the \ac{sem} could stop taking more frames and move on to the next imaging site. When the algorithm cannot make a confident prediction, this output could instruct the \ac{sem} to enhance the number of frames taken so a more confident prediction can be made. Overall, this can lead to a better balance between inspection performance and imaging throughput without requiring manual imaging recipe optimization. Other imaging parameters that could be modified dynamically for this purpose are pixel and field-of-view sizes. Algorithms that can make useful decisions about the imaging recipe faster than a few \ac{sem} imaging operations (e.g., taking a few frames) are needed to enable this. Likely contenders for such algorithms include combinations of \ac{dnn} feature extractors with bayesian optimization \cite{Kalinin2021} or reinforcement learning \cite{Kalinin2021, uzkent_zoom_2020}.

\subsubsection{Multi-modal \ac{dl}}\label{subsubsect:rq4_multimodal}
Few selected papers have used extra modal inputs to improve performance in defect inspection tasks. These extra modalities include energy-dispersive x-ray spectroscopy data \cite{sel_OLeary2020}, image Radon transformation \cite{sel_Yuan_Fu2020}, and pattern references \cite{sel_kim2021, sel_Ofir2022}. No selected \ac{dl} algorithms use \ac{mpsi} inputs, which have been shown to improve performance in \ac{nl} algorithms \cite{sel_Ben-Porath1999, sel_Tomlinson2000, sel_Hunt2000, sel_Abraham2001, sel_Serulnik2002, sel_Zontak2009, sel_Zontak2009b, sel_Lee2014}. \ac{dl} could enable easy and complete integration of many extra modalities for defect inspection due to its automatic feature extraction capabilities.

\section{Conclusion}\label{sect:conclusion}
In this paper, 103 research papers that propose automatic \ac{sem} image analysis algorithms for semiconductor defect inspection are reviewed. Each of these papers was first identified using search and filtering criteria from four popular publication databases. Secondly, the inspection tasks, algorithm types, and metrics used by these papers were identified and categorized. Thirdly, we answer four reasearch questions concerning (i) the common components of the algorithms, (ii) the effect of process context, (iii) the key challenges of current inspection algorithms, and (iv) promising directions for future work. In summary, \ac{dl} has become the most popular algorithm type, especially for classification tasks, with reference-based algorithms still being very relevant because of their complimentary strengths to \ac{dl}. We find that the type of algorithm for a particular task and use case should ultimately be decided based on which one can meet expected prediction performance with the fastest time-to-solution considering the available data.

\subsection*{Disclosures}
The authors have no conflicts of interest to disclose.

\subsection* {Code, Data, and Materials Availability} 
All data in support of the findings of this paper are available within the article or as supplementary material\cite{supplementary_material}.

\subsection*{Acknowledgments}
We thank the reviewers for their valuable feedback, which has significantly improved this review paper. Jesse Davis acknowledges the funding from the Flemish Government under the ``Onderzoeksprogramma Artifici\"ele Intelligentie (AI) Vlaanderen'' program. Microsoft 365 Copilot Chat was utilized to provide language and grammar correction suggestions throughout this manuscript. All outputs generated by the model were manually verified before being included in the final version.

\bibliography{references.bib}
\bibliographystyle{spiejour} 

\listoffigures
\listoftables

\vspace{2ex}\noindent\textbf{Enrique Dehaerne} is a doctoral student at the Declarative Languages and Artificial Intelligence (DTAI) section of the Department of Computer Science at KU Leuven. His current research, in collaboration with imec, focuses on deep learning for semiconductor pattern defect inspection. Enrique received the Master of Engineering: Computer Science degree, specializing in Artificial Intelligence, from KU Leuven, Belgium in 2022 and he was a Research Engineering Intern at Nokia Bell Labs, Antwerp in 2021. His research interests include computer vision, machine learning, semiconductor manufacturing and metrology, code generation, and robotics.

\vspace{2ex}\noindent\textbf{Bappaditya Dey} is a Senior R\&D Engineer, Advanced Patterning at imec, Belgium. In this role he develops deep learning methods for solving challenging industrial problems in EUV/EBEAM Lithography and SEM Metrology. Additionally, he is responsible for interdisciplinary research collaboration with multiple universities and research teams (across the globe) and mentoring students for their research thesis (MS/PhD) in this domain. He received his PhD degree, majoring in Computer Engineering, from the Center for Advanced Computer Studies (CACS), University of Louisiana at Lafayette, USA. He is a member of IEEE and SPIE. His research interests include VLSI, microelectronics, reconfigurable hardware, machine learning, computer vision, artificial intelligence, and semiconductor process optimization.

\vspace{2ex}\noindent\textbf{Victor M. Blanco Carballo} received his MSc in physics from university of Zaragoza followed by a PhD in semiconductor technology by University of Twente. He has been with ASML working as architect for defectivity and yield improvement and currently he is with imec working as team lead for patterning pathfinding. His current interests are focused on BEOL patterning and high NA development.

\vspace{2ex}\noindent\textbf{Jesse Davis} is a Professor in the Department of Computer Science at KU Leuven, Belgium. His research focuses on developing novel artificial intelligence, data science, and machine learning techniques, with a particular emphasis on analyzing structured data. Jesse's passions lie in using these techniques to make sense of lifestyle data, address problems in (elite) athlete monitoring and detect anomalies. Prior to joining KU Leuven, he obtained his bachelor's degree from Williams College, his PhD from the University of Wisconsin, and completed a post-doc at the University of Washington.

\end{spacing}
\end{document}